\documentclass{ptephy_v1}

\usepackage{pstricks}                                                                   
\usepackage{pstricks-add}
\begin{document}
\title{Scalar one-loop four-point Feynman integrals with complex internal masses}
\author{Khiem Hong Phan}
\affil{University of Science Ho Chi Minh City, 227 Nguyen Van Cu, Dist.5, Ho Chi Minh City, Vietnam
\email{phkhiem@hcmus.edu.vn}}
\begin{abstract}
Based on the method in Refs.~{\tt  [D.~Kreimer, Z.\ Phys.\ C {\bf 54} (1992) 667} and {\tt 
Int.\ J.\ Mod.\ Phys.\ A {\bf 8} (1993) 1797]}, we present analytic results for scalar
one-loop four-point Feynman integrals with complex internal masses. 
The results are not only valid for complex internal masses, but also for real internal mass cases. 
Different from the traditional approach proposed by G. 't Hooft and M. Veltman in the paper 
{\tt[Nucl.\ Phys.\ B {\bf 153} (1979) 365]}, this method can be extended to evaluate tensor integrals 
directly. Therefore, it may open a new approach to cure the inverse Gram determinant 
problem analytically. We then implement the results into a computer package which is 
{\tt ONELOOP4PT.CPP}. In numerical checks, one compares the program to {\tt LoopTools version}
$2.12$ in both real and complex mass cases. We find a perfect agreement 
between the results generated from this work and {\tt LoopTools}. 
\end{abstract}
\subjectindex{ One-loop Feynman integrals, numerical methods for quantum field theory.}
\maketitle
\section{Introduction}
\noindent
The future colliders, like the Large Hadron Collider (LHC) at high luminosities and the
International Linear Collider (ILC)~\cite{ATLAS:2013hta,CMS:2013xfa,Baer:2013cma}, 
aim to measure the properties of Higgs boson (or to explore the Higgs sector), of top quark 
and vector bosons, as well as search for Physics Beyond the Standard Model
(BSM). These measurements will be performed at high precision, e.g. the Higgs boson's couplings
will be measured at the precision of $1\%$ or better for a statistically significant 
measurement~\cite{Baer:2013cma}. In order to match the high precision data in the near
future, the higher-order corrections from theoretical calculations are necessary.
Therefore, the detailed evaluations for one-loop multi-leg and higher-loop at 
general scale to the selected scattering cross sections at the colliders are urgent 
requirements.

In traditional framework, the cross-section
of the processes at one-loop corrections will be obtained by integrating over the 
phase space of squared amplitudes which are decomposed into the tensor integrals. 
These tensor ones are then reduced into  scalar one-loop one-, two-, three- and four-point functions. 
It is well-known that the traditional tensor reductions may meet the inverse Gram determinant problem
\cite{Passarino:1978jh,Denner:2005nn} at several kinematic points in the phase space.
Consequently, it leads to numerical instabilities. One may apply the suitable 
experimental cuts to avoid the problem. However, the situation is completely
different when we consider one-loop multi-leg processes, for instance, $2\rightarrow 5,6,$ etc,
where the higher-point functions will be reduced to scalar one-loop one-, two-, three- and four-point 
functions with obtaining arbitrary configurations. As a result, we can not avoid the inverse Gram 
determinant problem as former case. Up to date the problem still has not been 
solved analytically and completely.

In the calculation of electroweak corrections to multi-particle processes, we have to 
handle  one-loop integrals with arbitrary internal mass and external momentum 
assignments. Moreover, for such processes involving unstable particles which can be on-shell, one 
has to resume their propagators by introducing complex masses~\cite{Denner:2005fg}. Therefore, 
the evaluations for tensor and scalar one-loop integrals at general scale, with complex 
internal masses, are also important.

The calculations for scalar one-loop one-, two-, three- and four-point functions are important 
ingredients for evaluating higher-order corrections. Pioneering calculation for these functions 
has been performed by 't~Hooft and Veltman~\cite{'tHooft:1978xw}. For the one-, two-, and 
three-point functions, the authors of Ref.~\cite{'tHooft:1978xw} have provided compact explicit expressions
that are valid for real and complex internal masses as well as at general scale. For the scalar four-point 
functions, an analytic result for real mass cases has been presented in 
Ref.~\cite{'tHooft:1978xw,Denner:1991qq}. However, the complex mass cases 
have been only discussed in these papers. Recently, Ref.~\cite{Nhung:2009pm} 
has been extended the method of 't~Hooft and Veltman for computing 
scalar one-loop four-point functions with 
complex internal masses. It has been already implemented into a 
{\sc FORTRAN} program, named {\tt D$0$C}.  
In addition, the authors of Ref.~\cite{Denner:2010tr}
has followed the previous works in Refs.~\cite{'tHooft:1978xw,Denner:1991qq} for 
evaluating these functions with complex internal 
masses in which all infrared divergence (IR) cases have been treated completely.

Together with these works, it is worth to mention the references
\cite{vanOldenborgh:1989wn,Ellis:2007qk,vanHameren:2010cp,Binoth:2008uq,Cullen:2011kv, 
Guillet:2013msa}. In general statements, these calculations for scalar one-loop Feynman integrals 
have been followed the traditional approach developed by 't~Hooft and Veltman.  
Alternative methods we mention in this paper are~\cite{Duplancic:2000sk,Bern:1992em}. 
In these papers, the methods have been only developed for evaluating one-loop 
corrections to QCD processes.

Based on above methods, there are many computer packages which
are available such as~\cite{vanOldenborgh:1989wn,Hahn:1998yk,
Berger:2008ag, Ossola:2007ax, Carrazza:2016gav, Actis:2016mpe, Denner:2016kdg}. 
To our knowledge, these calculations and packages still have not cured the inverse 
Gram determinant problem analytically and completely. Besides that, some of them 
has still not provided scalar one-loop integrals for complex masses.

Direct Computation Method (DCM), which is a purely numerical method, has been 
applied to evaluate Feynman integrals. The method is based on the combination 
of an efficient numerical integration and extrapolation~\cite{Yuasa:2011kt}. 
Scalar one-loop integrals for real/complex masses have been calculated 
successfully in this approach.  Another purely numerical program, which is 
SecDec~\cite{Heinrich:2008si, Borowka:2012yc, Borowka:2015mxa,Borowka:2017idc}, 
is based on the sector decomposition method. Numerical Mellin-Barnes 
representations for Feynman integrals also have been presented in 
Ref.~\cite{Gluza:2007rt}.

Solving Gram determinant problem analytically and proving an 
alternative method for evaluating scalar one-loop Feynman integrals, special for four-point functions,  
with including complex internal masses, are mandatory. In the scope of this paper, based on 
the method developed in Refs.~\cite{Kreimer:1991wj, Kreimer:1992ps, Franzkowski}, we present 
analytic results for scalar one-loop four-point Feynman integrals
with complex internal masses. The results are not only valid for complex 
internal masses, but also for real mass cases. Different from the method proposed by G. 't~Hooft and 
M. Veltman in the paper~\cite{'tHooft:1978xw}, this method can be extended to evaluate tensor 
integrals directly. Therefore, it may open a new approach to cure the inverse Gram determinant 
analytically. We then implement the results into a computer package which is {\tt ONELOOP4PT.CPP}. 
In numerical checks, one compares this work to 
{\tt LoopTools version $2.12$}~\cite{Hahn:1998yk} in both real and complex mass cases. 
We find a perfect agreement between the results generated from this work and {\tt LoopTools}.

The layout of the paper is as follows: In section 2,  we present the method for evaluating scalar one-loop 
four-point functions in detail. In section 3, we will show the numerical checks on 
the program with {\tt LoopTools}. Conclusions and plans for future work are presented in section 4.  
Several useful formulae used in this calculation are shown in the Appendix.
\section{The calculation} 
\begin{figure}[ht]
\begin{center}
\begin{pspicture}(-6, -3)(6, 3)
\psline(-2,-2)(-2,2)
\psline{->}(-2, -2)(-2, 0)
\psline(2,-2)(2,2)
\psline{->}(2, 2)(2, 0)
 \psline(-2,2)(2,2)
\psline{->}(-2, 2)(0, 2)
\psline(-2,-2)(2,-2)
\psline{->}(2, -2)(0, -2)
\psline(-4,-2)(-2,-2)
\psline{->}(-4,-2)(-3,-2)
\psline(-4,2)(-2,2)
\psline{->}(-4,2)(-3,2)
\psline(2,-2)(4,-2)
\psline{->}(4,-2)(3,-2)
\psline(2,2)(4,2)
\psline{->}(4,2)(3,2)
\rput(-3,2.5){$ p_2$}
\rput(-3,-2.5){$ p_1$}
\rput(3,2.5){$ p_3$}
\rput(3,-2.5){$ p_4$}
\rput(0,-2.5){$ l$}
\rput(0,1.5){$ l+q_2$}
\rput(-3,0){$ l+q_1$}
\rput(3,0){$ l+q_3$}
\psset{dotsize=4pt 0}
\psdots(-2, -2)(2, 2)(-2, 2)(2,- 2)
\end{pspicture}
\caption{\label{d0feyn} The box diagrams.}
\end{center}
\end{figure}
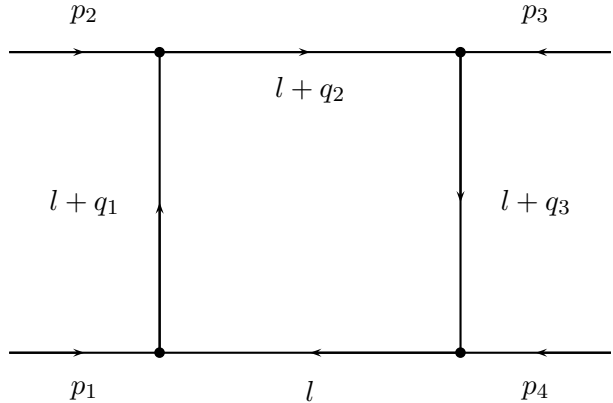

In this section, we apply the method which was proposed in Refs.~\cite{Kreimer:1991wj, Kreimer:1992ps, Franzkowski} to calculate scalar one-loop four-point functions. The Feynman integrals for these functions are given by
\begin{eqnarray}
\label{d0}
D_0\equiv 
D_0(p_1^2, p_2^2, p_3^2, p_4^2, s, t, m_1^2, m_2^2, m_3^2, m_4^2) 
:= \int\dfrac{d^nl}{P_1 P_2 P_3 P_4}.
\end{eqnarray}
Where the inverse Feynman propagators are  
\begin{eqnarray}
P_1 &=& (l+p_1)^2 - m_1^2+i\rho,                  \\
P_2 &=& (l+p_1+p_2)^2-m_2^2+i\rho,            \\
P_3 &=& (l+p_1+p_2+p_3)^2-m_3^2 +i\rho,   \\
P_4 &=& l^2- m_4^2+i\rho.                             
\end{eqnarray}
The term $i\rho$ is Feynman's prescription. $p_i$ and  $m_i$  for $i =1,2,3,4$ 
are external momenta and internal masses respectively. The external momenta
flow incoming as Fig.~\ref{d0feyn} and follow momentum conservation law 
$p_1 +p_2 +p_3+ p_4=0$. The loop momentum is $l$ and $n$ is space-time dimension.
In this calculation, we are not going to deal with infrared divergence. Thus, 
we  will work directly in space-time dimension $n=4$. In general, $D_0$ is a function of 
$p_1^2, p_2^2, p_3^2, p_4^2, s, t, m_1^2, m_2^2, m_3^2, m_4^2$ 
with $s= (p_1+p_2)^2$, $t= (p_2+p_3)^2$. 

In parallel and orthogonal space \cite{Kreimer:1991wj, Kreimer:1992ps}
which is spanned by external momenta, the Feynman integral is obtained 
\begin{eqnarray}
D_0&=&2\int\limits_{-\infty}^{\infty} \;
d l_{0}dl_{1} dl_{2}\int\limits_{0}^{\infty}d l_{\bot}\dfrac{1}{P_1P_2P_3P_4}.
\end{eqnarray}
Let us define the momenta $q_i =\sum\limits_{j=1}^i p_j$ for
$i, j=1,2, \cdots,4$. One arrives at the case of at least one 
time-like momentum, $p_i^2>0$ for $i=1,2,\cdots, 4$, 
e.g. $p_1^2>0$.  Working in the rest frame of $p_1$, one has
\begin{eqnarray}
\label{Lorentz}
q_{1}&=&(q_{10},0,0,0),                               \\
q_{2}&=& (q_{20},q_{21},0,0),          \\
q_{3}&=& (q_{30},q_{31},q_{32},0),    \\
q_{4}&=&0.    
\end{eqnarray}
In the complex mass scheme, the internal masses are taken the form 
\begin{eqnarray}
 m_k^2&=&m_{0k}^2-i m_{0k} \Gamma_{k},
\end{eqnarray}
 $\Gamma_{k}\geqslant 0$ are decay widths of unstable particles 
 $k = 1,2, \cdots, 4$. \\
 
The Feynman integral is then written explicitly as
\begin{eqnarray}
D_0&=&2\int\limits_{-\infty}^{\infty} d l_{0}dl_{1} dl_{2}\int\limits_{0}^{\infty}d l_{\bot} \times \\
&&\times \dfrac{1}{[(l_0+q_{10})^2-l_1^2-l_2^2-l_{\perp}^2-m_1^2+i\rho]
[(l_0+q_{20})^2-(l_1+q_{21})^2-l_2^2-l_{\perp}^2-m_2^2+i\rho]}\nonumber\\
&&\times \dfrac{1}{[(l_0+q_{30})^2-(l_1+q_{31})^2-(l_2+q_{32})^2-l_{\perp}^2-m_3^2+i\rho]\
[l_0^2-l_1^2-l_2^2-l_{\perp}^2-m_4^2+i\rho]}.\nonumber
\end{eqnarray}
 
Partitioning the integrand into the form
\begin{eqnarray}
\dfrac{1}{P_1P_2P_3P_4}&=&\frac{1}{P_1(P_2-P_1)(P_3-P_1)(P_4-P_1)}
+ \frac{1}{P_2(P_1-P_2)(P_3-P_2)(P_4-P_2)}\nonumber\\
& &+\frac{1}{P_3(P_1-P_3)(P_2-P_3)(P_4-P_3)}+\frac{1}{P_4(P_1-P_4)(P_2-P_4)(P_3-P_4)}\nonumber\\
&=&\sum_{k=1}^{4}\dfrac{1}{P_k\prod\limits_{\substack{l=1\\l\neq k}}(P_{l}-P_{k})},
\end{eqnarray}
we then make a shift on $P_k$ by $l_i\longrightarrow l_i+q_{ki}$ for $k=1,2,\cdots,4$ and $i=0,1,\cdots, 3$, 
the resulting reads
\begin{eqnarray}
\label{D0afterpar}
D_0&=&2\sum_{k=1}^{4}\int\limits_{-\infty}^{\infty}dl_0dl_1dl_2\int\limits_{0}^{\infty}dl_{\perp}\times 
\nonumber\\
&&\times \dfrac{1}{\Big[l_0^2-l_1^2-l_2^2-l_{\perp}^2-m_k^2+i\rho \Big]}
\dfrac{1}{\prod\limits_{\substack{l=1\\k\neq l}}^{4}\Big[a_{lk}l_0+b_{lk}l_1+c_{lk}l_2+d_{lk}\Big]}.  
\end{eqnarray}
We have already introduced the  following kinematic variables 
\begin{equation}
 \begin{array}{ll}
a_{lk} = 2(q_{l0} -q_{k0}), & b_{lk} = - 2(q_{l1} -q_{k1}),\\
c_{lk} = - 2(q_{l2} -q_{k2}), & d_{lk} = (q_l -q_k)^2 -(m_l^2 -m_k^2).
\end{array}
\end{equation}
It is important to note that the kinematic variables $a_{lk}, b_{lk}, c_{lk} \in \mathbb{R}$ 
and $ d_{lk}\in \mathbb{C}$.  In the next subsections, we are going to calculate this integral.
\subsection{Linearization and $l_0$-integration} 
To integrate over $l_0$, we first linearize $l_0$  by applying 
a transform as
\begin{eqnarray}
\begin{array}{ll}
l_0=x+z, &l_1 =y,\\
l_2=x, &l_{\perp}=t,
\end{array}
\end{eqnarray}
then the Jacobian of this transformation is
$ |J|=\Big|\dfrac{\partial(l_0,l_1,l_2,l_{\perp})}{\partial(z,y,x,t)}\Big|=1$. 
We arrive at
\begin{eqnarray}
\label{residumx1}
 D_0&=& 2\sum_{k=1}^{4}\int\limits_{-\infty}^{\infty}dxdy
 \int\limits_{-\infty}^{\infty}dz\int\limits_{0}^{\infty}dt \;\mathcal{F}(x,y,z,t),
\end{eqnarray}
with the integrand is corresponding to 
\begin{eqnarray}
\mathcal{F}(x,y,z,t)&=&\dfrac{1}{\left[2xz-z^2-y^2-t^2-m_k^2+i \rho \right]
\prod\limits_{\substack{l=1\\k\neq l}}^{4}\left[ a_{lk}\; z+b_{lk}\; y+AC_{lk}\; x 
+ d_{lk} \right] }.
\end{eqnarray}
In this integrand, we have already used new variables
\begin{eqnarray}
 AC_{lk}=a_{lk}+c_{lk} \in \mathbb{R}. 
\end{eqnarray}

The integrand now depends linearly on $x$. Thus,  the $x$-integration will be taken 
easily by applying the residue theorem. For this purpose, one should first 
analyze the $x$-poles of this integrand. These poles are 
\begin{eqnarray}
\label{xpoles}
 x_l&=& -\Big[\frac{a_{lk}}{AC_{lk}}z +\frac{b_{lk}}{AC_{lk}}y+\frac{d_{lk}}{AC_{lk}}\Big], \\
 && \nonumber\\
x_0 &=& \dfrac{z^2+y^2+t^2+m_k^2-i \rho}{2z}.  
\end{eqnarray}
We realize that $\mathrm{Im}\left( \dfrac{z^2+y^2+t^2+m_k^2-i \rho}{2z} \right) 
=- \dfrac{m_{0k} \Gamma_{k} +\rho}{2z}$ which depends on the sign of $z$.  The location of 
$x_0$ in the $x$-complex plane will be determined by the sign of $z$, see Fig.~\ref{zneg} for 
more detail. We therefore should rewrite $D_0$ as following:
\begin{eqnarray}
D_0=D_0^{+}+D_0^{-},
\end{eqnarray}
with
\begin{eqnarray}
D_0^{+} &=& 
2\sum_{k=1}^{4}\int\limits_{-\infty}^{\infty}dxdy\int\limits_{0}^{\infty}dz\int\limits_{0}^{\infty}dt  \;\mathcal{F}(x,y,z,t),\\
D_0^{-}   &=& 
2\sum_{k=1}^{4}\int\limits_{-\infty}^{\infty}dxdy\int\limits^{0}_{-\infty}dz\int\limits_{0}^{\infty}dt  \;\mathcal{F}(x,y,z,t).
\end{eqnarray}
\vspace*{-1.25cm}
\begin{figure}[ht]
\begin{tabular}{cc} 
\begin{pspicture}(-1, -2)(6, 5)
\label{zneg}
\psarc{->}(3, 1){2.5}{0}{92}
\psarc(3, 1){2.5}{92}{180}
\psline(5.5,1)(4.7, 1)
\psline(4.3, 1)(3.2, 1)
\psline(2.8, 1)(1.7, 1)
\psline(1.3, 1)(0.5,1)
\psarc{->}(4.5, 1){0.2}{180}{272}
\psarc(4.5, 1){0.2}{272}{360}
\psarc{->}(3, 1){0.2}{180}{272}
\psarc(3, 1){0.2}{272}{360}
\psarc{->}(1.5, 1){0.2}{180}{272}
\psarc(1.5, 1){0.2}{272}{360}
\psline(6,1)(5.5, 1)
\psline(0.5,1)(0, 1)
\rput(4.5,1){$\otimes$}
\rput(3,1){$\otimes$}
\rput(1.5,1){$\otimes$}
\rput(5,0){$\otimes$}
\rput(3, 1.8){$ z > 0$}
\end{pspicture}
&
\begin{pspicture}(-8, 0)(6, 5)
\label{zpos}
\psarc(-3, 3){2.5}{180}{272}
\psarc{<-}(-3, 3){2.5}{272}{360}
\psline(-5.5,3)(-4.7, 3)
\psline(-4.3, 3)(-3.2, 3)
\psline(-2.8, 3)(-1.7, 3)
\psline(-1.3, 3)(-0.5,3)
\psarc{<-}(-4.5, 3){0.2}{80}{180}
\psarc(-4.5, 3){0.2}{0}{80}
\psarc{<-}(-3, 3){0.2}{80}{180}
\psarc(-3, 3){0.2}{0}{80}
\psarc{<-}(-1.5, 3){0.2}{80}{180}
\psarc(-1.5, 3){0.2}{0}{80}
\psline(-6,3)(-5.5, 3)
\psline(-0.5,3)(0, 3)
\rput(-4.5,3){$\otimes$}
\rput(-3,3){$\otimes$}
\rput(-1.5,3){$\otimes$}
\rput(-5,4){$\otimes$}
\rput(-3, 1.8){$ z < 0$}
\end{pspicture}
\end{tabular}
\caption{\label{zneg} The contour integration in the $x$-plane.}
\end{figure}
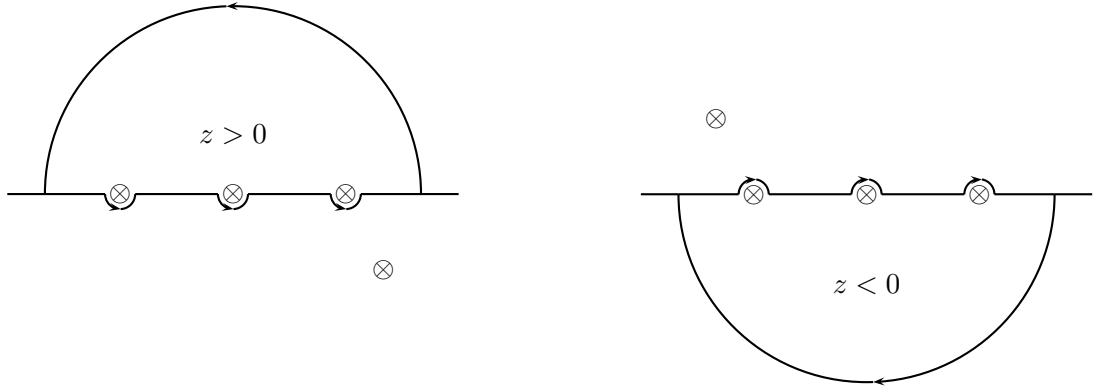

By closing the integration contour over upper $x$ plane  when
$z>0$ and vise versa, as shown in Fig.~\ref{zneg}, one then applies the
residue theorem, the resulting then reads
 \begin{eqnarray}
\label{Doplus}
 D_0^{+}&=&2\pi i\sum_{k=1}^{4}\sum_{\substack{l=1\\ l\neq k}}^{4}\dfrac{1}{AC_{lk}}
 \int\limits_{-\infty}^{\infty}dy\int\limits_{0}^{\infty}dz\int\limits_{0}^{\infty}dt\;
 \dfrac{1}{\prod\limits_{\substack{m=1\\m\neq l\\l\neq k}}(A_{mlk}z+B_{mlk}y+C_{mlk})}\nonumber\\
&&\hspace{1.2cm}\times \;\dfrac{f_{lk}\Big(1-\delta(AC_{lk})\Big)}{\Big[\Big(1-\frac{2a_{lk}}{AC_{lk}}\Big)z^2
-\frac{2b_{lk}}{AC_{lk}}yz-\frac{2d_{lk}}{AC_{lk}}z-y^2-t^2-m_k^2+i \rho \Big]},
\end{eqnarray}
and 
\begin{eqnarray}
\label{Dominus}
D_0^{-}&=&-2\pi i\sum_{k=1}^{4}\sum_{\substack{l=1\\l\neq k}}^{4}\frac{1}{AC_{lk}}
\int\limits_{-\infty}^{\infty}dy\int\limits_{-\infty}^{0}dz\int\limits_{0}^{\infty}dt
\;\dfrac{1}{\prod\limits_{\substack{m=1\\m\neq l,k}}(A_{mlk}z+B_{mlk}y+C_{mlk})}\nonumber\\
&&\hspace{1.7cm}\times \;\dfrac{f^{-}_{lk}\Big(1-\delta(AC_{lk})\Big)}{\Big[\Big(1-\frac{2a_{lk}}{AC_{lk}}\Big)z^2
-\frac{2b_{lk}}{AC_{lk}}yz-\frac{2d_{lk}}{AC_{lk}}z-y^2-t^2-m_k^2+i \rho \Big]}. 
\end{eqnarray}
The $ f_{lk}$ and $  f_{lk}^{-}$ functions indicate the residue contributions
from the $x$-poles in Eq.~(\ref{xpoles}). These functions 
are defined as
\begin{eqnarray}
\label{flk}
f_{lk}=
\begin{cases}
          0,     & \text{if}   \;\; \text{Im}\left(-\dfrac{d_{lk}}{AC_{lk}}\right)<0;\\
          1,     & \text{if}   \;\;\text{Im}\left(-\dfrac{d_{lk}}{AC_{lk}}\right)=0;\\
          2,     & \text{if}   \;\; \text{Im}\left(-\dfrac{d_{lk}}{AC_{lk}}\right)>0.
 \end{cases}
\quad \text{and} \quad 
f_{lk}^{-}=
 \begin{cases}
0,   &\text{if}    \;\;  \text{Im}\left(-\dfrac{d_{lk}}{AC_{lk}} \right)>0;\\
1,   & \text{if}    \;\; \text{Im}\left(-\dfrac{d_{lk}}{AC_{lk}} \right)=0;\\
2,   &  \text{if}   \;\; \text{Im}\left(-\dfrac{d_{lk}}{AC_{lk}} \right)<0.
\end{cases}
\end{eqnarray}

The new kinematic variables introduced in this step 
are listed as following
\begin{eqnarray}
A_{mlk}&=&a_{mk}-\frac{a_{lk}}{AC_{lk}}AC_{mk}, \\
B_{mlk}&=&b_{mk}-\frac{b_{lk}}{AC_{lk}}AC_{mk},\\
C_{mlk}&=&d_{mk}-\frac{d_{lk}}{AC_{lk}}AC_{mk}.
\end{eqnarray}
The delta function is defined 
\begin{equation}
 \delta(AC_{lk})=
 \begin{cases}
        0,   &\text{if}    \;\;  AC_{lk} \neq 0;\\
        1,   & \text{if}    \;\;AC_{lk}=0.
\end{cases}
\end{equation}
It is important to note that  $A_{mlk}, B_{mlk} \in \mathbb{R}$, and $C_{mlk}\in \mathbb{C}$.
Combining with the definition of $f_{lk}$ and $ f_{lk}^{-}$ in Eq.~(\ref{flk}), 
we verify easily that 
\begin{eqnarray}
\label{Imflk}
\mathrm{Im} 
\Big[ \Big(1-\frac{2a_{lk}}{AC_{lk}}\Big)z^2-\frac{2b_{lk}}{AC_{lk}}yz
-\frac{2d_{lk}}{AC_{lk}}z-y^2-t^2-m_k^2+i \rho   \Big] > 0.
\end{eqnarray}
\subsection{The $y$-integration} 
We are now going to evaluate the three-fold integrals which arrived the previous subsection, see 
Eqs.~(\ref{Doplus}, \ref{Dominus}). In order to work out the $y$-integration, one has  to 
linearize $y$ by using the Euler shift $t\rightarrow t+y$.  However, we realize that the terms 
proportional to $t^2$ and $y^2$ in the integrand have the same sign. 
One can first make $t^2$ and $y^2$ having opposite sign by applying a complex rotation 
in the $t$-plane. 

\begin{figure}[ht]
\begin{center}
\begin{pspicture}(-6, -3)(6, 3)
\psline{->}(0,-3)(0,3)
\psline{->}(-5,0)(5,0)
\psline{->}(0.2,-2.8)(0.2,-1.3)
\psline(0.2,-1.3)(0.2,-0.2)
\psline{->}(0.2,-0.2)(2,-0.2)
\psline(2,-0.2)(4,-0.2)
\psbezier(0.2,-2.8)(1,-2.7)(3,-2.5)(4,-0.2)
\rput(0.7,3){$\mathrm{Im}(t)$}
\rput(4.5,0.5){$\mathrm{Re}(t)$}
\rput(3,1){$\otimes$}
\rput(-3,-1){$\otimes$}
\rput(3,-2){$C_{k}$}
\end{pspicture}
\end{center}
\caption{\label{wickkk} $t-$rotation}
\end{figure}
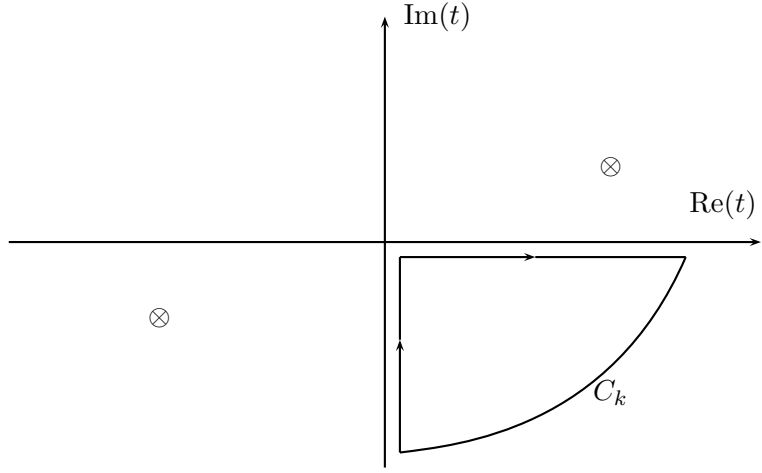

The integrand written in terms of $t$ have two poles which are
\begin{eqnarray}
 t_{1,2}=\pm\sqrt{\Big(1-\frac{2a_{lk}}{AC_{lk}}\Big)z^2-\frac{2b_{lk}}{AC_{lk}}yz 
 - \frac{2d_{lk}}{AC_{lk}}z-y^2-m_k^2+i \rho}. 
\end{eqnarray}
Because of Eq.~(\ref{Imflk}), we find that $t_{1,2}$ are located in the first and the third 
quarters of $t$-complex plane, as described in  Fig.~\ref{wickkk}.  As a matter of this fact, one should 
choose the integration contour on the fourth quarter of the $t$-complex plane, as Fig.~\ref{wickkk}. 
There are no residua of $t$-poles contribute to the $t$-integration contour.  
We, therefore, derive the following relation
\begin{eqnarray}
\label{wickform}
\int\limits_{0}^{\infty}dt=\frac{1}{2}\int\limits_{-\infty}^{\infty}dt= 
-\int\limits^{0}_{-i\infty}dt
=-\frac{1}{2}\int\limits_{-i\infty}^{i\infty}dt.
\end{eqnarray}
We are now applying the relation in Eq.~(\ref{wickform}). One then 
makes the rotation like $t\longrightarrow it$. 
The resulting reads 
\begin{eqnarray}
D_0^{+}&=&\pi\sum_{k=1}^{4}\sum_{\substack{l=1\\l\neq k}}^{4}\frac{1}{AC_{lk}}
\int\limits_{-\infty}^{\infty}dy\int\limits_{0}^{\infty}dz\int\limits_{-\infty}^{\infty}dt\;
\dfrac{1}{\prod\limits_{\substack{m=1\\m\neq l,k}}(A_{mlk}z+B_{mlk}y+C_{mlk})}\\
&&\hspace{1.5cm} \;\dfrac{f_{lk}\Big(1-\delta(AC_{lk})\Big)}{\Big[\Big(1-\frac{2a_{lk}}{AC_{lk}}\Big)z^2
-\frac{2b_{lk}}{AC_{lk}}yz-\frac{2d_{lk}}{AC_{lk}}z-y^2 + t^2-m_k^2+i \rho \Big]}, 
\nonumber\\
D_0^{-}&=&-\pi\sum_{k=1}^{4}\sum_{\substack{l=1\\l\neq k}}^{4}\frac{1}{AC_{lk}}
\int\limits_{-\infty}^{\infty}dy\int\limits^{0}_{-\infty}dz\int\limits_{-\infty}^{\infty}dt
\;\dfrac{1}{\prod\limits_{\substack{m=1\\m\neq l,k}}(A_{mlk}z+B_{mlk}y+C_{mlk})}\\
&&\hspace{2cm} \;\dfrac{f^{-}_{lk}\Big(1-\delta(AC_{lk})\Big)}{\Big[\Big(1-\frac{2a_{lk}}{AC_{lk}}\Big)z^2 
- \frac{2b_{lk}}{AC_{lk}}yz-\frac{2d_{lk}}{AC_{lk}}z-y^2 + t^2-m_k^2+i \rho \Big]}.
\nonumber
\end{eqnarray}
After obtaining the opposite sign of $t^2$and $y^2$ in these integrands, we proceed the 
linearization of $y$ by performing the above Euler transformation, 
or $t\rightarrow t+y$. We arrive at
\begin{eqnarray}
\label{D0afterwick1}
D_0^{+}&=&+\pi\sum_{k=1}^{4}\sum_{\substack{l=1\\l\neq k}}^{4}\frac{1}{AC_{lk}}
\int\limits_{-\infty}^{\infty}dy\int\limits_{0}^{\infty}dz\int\limits_{-\infty}^{\infty}dt\;
\dfrac{1}{\prod\limits_{\substack{m=1\\m\neq l,k}}(A_{mlk}z+B_{mlk}y+C_{mlk})}\\
&&\hspace{1cm}\times \;\dfrac{f_{lk}\Big(1-\delta(AC_{lk})\Big)}{\Big[\Big(1-\frac{2a_{lk}}{AC_{lk}}\Big)z^2
+2\Big(t-\frac{b_{lk}}{AC_{lk}}z\Big)y-\frac{2d_{lk}}{AC_{lk}}z+t^2-m_k^2+i \rho \Big]}, 
\nonumber
\end{eqnarray}
\begin{eqnarray}
\label{D0afterwick2}
D_0^{-}&=&-\pi\sum_{k=1}^{4}\sum_{\substack{l=1\\l\neq k}}^{4}\frac{1}{AC_{lk}}
\int\limits_{-\infty}^{\infty}dy\int\limits^{0}_{-\infty}dz\int\limits_{-\infty}^{\infty}dt
\;\dfrac{1}{\prod\limits_{\substack{m=1\\m\neq l,k}}(A_{mlk}z+B_{mlk}y+C_{mlk})}\\
&&\hspace{1cm}\times \;\dfrac{f^{-}_{lk}\Big(1-\delta(AC_{lk})\Big)}{\Big[\Big(1-\frac{2a_{lk}}{AC_{lk}}\Big)z^2
+ 2\Big(t-\frac{b_{lk}}{AC_{lk}}z\Big)y-\frac{2d_{lk}}{AC_{lk}}z+t^2-m_k^2+i \rho \Big]}. \nonumber
\end{eqnarray}

The $y$-integration will be taken by using the residue theorem. 
The locations of $y$-poles are more complicated than in the case of $x$-integration. 
According to Eqs.~(\ref{D0afterwick1},\ref{D0afterwick2}), combining with Eq.~(\ref{Imflk}),
it is easy to check that the imaginary parts of the $y$-poles in these integrands are
\begin{eqnarray}
\text{Im}\left(\frac{\frac{2d_{lk}}{AC_{lk}}z + m_k^2-i \rho}{t-\frac{b_{lk}}{AC_{lk}}z } \right),
\end{eqnarray}
which depend on the sign of $t-\frac{b_{lk}}{AC_{lk}}z$. Similar to the $x$-integration, we should 
cut the integration of $t$ into two segments $t \geqslant \alpha_{lk}z$ and $t\leqslant \alpha_{lk}z$ with 
$\alpha_{lk}=\frac{b_{lk}}{AC_{lk}} $.  The imaginary parts of the remaining poles which 
are the roots of the following equation
\begin{eqnarray}
\label{yl}
 A_{mlk}z+B_{mlk}y+C_{mlk} =0,
\end{eqnarray}
become more complicated. They then contribute to the residua of the taken 
integrations.

Closing the contour on the upper plane of $y$ if $t \geqslant \alpha_{lk}z$ and vice versa,  one takes into account
the residua of the $y$-poles in Eq.~(\ref{yl}). Finally, we arrive at
\begin{eqnarray}
 D_0 &=& D_0^{++}+D_0^{+-}+D_0^{-+}+D_0^{--},
\end{eqnarray}
with
\begin{eqnarray}
D_0^{++}&=&+i\pi^2\sum_{\substack{m,l,k=1\\l\neq k\neq m}}^{4}\frac{\Big(1-\delta(AC_{lk})\Big)
\Big(1-\delta(B_{mlk})\Big)}{AC_{lk}(A_{nlk}B_{mlk}-A_{mlk}B_{nlk})}
\int\limits_{0}^{\infty}dz\int\limits_{\alpha_{lk}z}^{\infty}dt\; f_{lk}g_{mlk} \; \mathcal{I}'_{nmlk}(z,t),\nonumber\\
&& \\
D_0^{+-}&=&-i\pi^2\sum_{\substack{m,l,k=1\\l\neq k\neq m}}^{4}\frac{\Big(1-\delta(AC_{lk})\Big)
\Big(1-\delta(B_{mlk})\Big)}{AC_{lk}(A_{nlk}B_{mlk}-A_{mlk}B_{nlk})}
\int\limits_{0}^{\infty}dz\int\limits^{\alpha_{lk}z}_{-\infty}dt\; f_{lk}g_{mlk}^{-} \; \mathcal{I}'_{nmlk}(z,t),\nonumber\\
&& \\
D_0^{-+}&=&-i\pi^2\sum_{\substack{m,l,k=1\\l\neq k\neq m}}^{4}\frac{\Big(1-\delta(AC_{lk})\Big)
\Big(1-\delta(B_{mlk})\Big)}{AC_{lk}(A_{nlk}B_{mlk}-A_{mlk}B_{nlk})}
\int\limits^{0}_{-\infty}dz\int\limits_{\alpha_{lk}z}^{\infty}dt\; f_{lk}^{-}g_{mlk}  \; \mathcal{I}'_{nmlk}(z,t),\nonumber\\
&& \\
D_0^{--}&=&+i\pi^2\sum_{\substack{m,l,k=1\\l\neq k\neq m}}^{4}\frac{\Big(1-\delta(AC_{lk})\Big)
\Big(1-\delta(B_{mlk})\Big)}{AC_{lk}(A_{nlk}B_{mlk}-A_{mlk}B_{nlk})}
\int\limits^{0}_{-\infty}dz\int\limits^{\alpha_{lk}z}_{-\infty}dt\; f_{lk}^{-}g_{mlk}^{-} \; \mathcal{I}'_{nmlk}(z,t),\nonumber\\
\end{eqnarray}
and the integrand 
\begin{eqnarray}
 I'_{nmlk}(z,t)&=&\dfrac{1}{\Big[z+F_{nmlk}\Big]\Big[D'_{mlk}z^2-2\frac{A_{mlk}}{B_{mlk}}zt - 
 2\frac{C_{mlk}}{B_{mlk}}t +E'_{mlk}z+t^2-m_k^2+i\rho \Big]}. \nonumber\\
 \end{eqnarray}
We have already introduced following kinematic variables: 
 \begin{eqnarray}
F_{nmlk}&=&\frac{C_{nlk}B_{mlk}-B_{nlk}C_{mlk}}{A_{nlk}B_{mlk}-B_{nlk}A_{mlk}},\\
D'_{mlk}&=&1-\frac{2a_{lk}}{AC_{lk}}+ 2\frac{b_{lk}}{AC_{lk}}\frac{A_{mlk}}{B_{mlk}},\\
E'_{mlk}&=&-2\Big( \frac{d_{lk}}{AC_{lk}} -\frac{b_{lk}}{AC_{lk}}\frac{C_{mlk}}{B_{mlk}}\Big).
\end{eqnarray}
The $g_{mlk}$ and $g^-_{mlk}$ functions will indicate the locations of 
$y$-poles in Eq.~(\ref{yl}) which contributed to the integrations.
They are defined as
\begin{eqnarray}
\label{gmlk}
g_{mlk}= 
\begin{cases}
          0,     & \text{if}   \;\; \mathrm{Im} \left(-\dfrac{C_{mlk}}{B_{mlk}} \right)<0;\\
          1,     & \text{if}   \;\; \mathrm{Im}\left(-\dfrac{C_{mlk}}{B_{mlk}} \right)=0;\\
          2,     & \text{if}   \;\; \mathrm{Im}\left(-\dfrac{C_{mlk}}{B_{mlk}}\right)>0;
 \end{cases} 
\qquad  \text{and}  \qquad
g_{mlk}^{-}=
\begin{cases}
          0,     & \text{if}   \;\; \mathrm{Im}\left (-\dfrac{C_{mlk}}{B_{mlk}} \right)>0;\\
          1,     & \text{if}   \;\; \mathrm{Im}\left(-\dfrac{C_{mlk}}{B_{mlk}} \right)=0;\\
          2,     & \text{if}   \;\; \mathrm{Im}\left (-\dfrac{C_{mlk}}{B_{mlk}} \right)<0.
 \end{cases}
\end{eqnarray}
We now make a shift $t\longrightarrow t'=t-\alpha_{lk}z$. The Jacobian of this shift is $1$
and the $t$-integrals change the border to $[0, \pm \infty]$. The resulting reads
\begin{eqnarray}
\label{yint1}
D_0^{++}&=&+i\pi^2\bigoplus_{nmlk}\int\limits_{0}^{\infty}dz\int\limits_{0}^{\infty}dt\;\;\; f_{lk}g_{mlk} \;\; \mathcal{I}_{nmlk}(z,t),\\
\label{yint2}
D_0^{+-}&=&-i\pi^2\bigoplus_{nmlk}\int\limits_{0}^{\infty}dz\int\limits^{0}_{-\infty}dt\;\;\; f_{lk}g_{mlk}^{-} \;\;  \mathcal{I}_{nmlk}(z,t),\\
\label{yint3}
D_0^{-+}&=&-i\pi^2\bigoplus_{nmlk}\int\limits^{0}_{-\infty}dz\int\limits_{0}^{\infty}dt\;\;\; f_{lk}^{-}g_{mlk} \;\;  \mathcal{I}_{nmlk}(z,t),\\
\label{yint4}
D_0^{--}&=&+i\pi^2\bigoplus_{nmlk}\int\limits^{0}_{-\infty}dz\int\limits^{0}_{-\infty}dt\;\;\; f_{lk}^{-}g_{mlk}^{-}\;\;  \mathcal{I}_{nmlk}(z,t),
\end{eqnarray}
with the new notations 
\begin{eqnarray}
\label{integrandz}
\bigoplus_{nmlk}&=&\sum_{\substack{k=1}}^{4}\sum_{\substack{l=1\\l\neq k}}^{4}
\sum_{\substack{m=1\\m\neq l \\m\neq k}}^{4}\frac{1}{AC_{lk}(A_{nlk}B_{mlk}-A_{mlk}B_{nlk})}
\Big[1-\delta(AC_{lk})\Big]\Big[1-\delta(B_{mlk})\Big], 
\\
\mathcal{I}_{nmlk}(z,t)&=&\dfrac{1}{\Big[z+F_{nmlk}\Big]\Big[D_{mlk}z^2 
- 2\Big(\frac{A_{mlk}}{B_{mlk}}-\alpha_{lk}\Big)zt-2\frac{C_{mlk}}{B_{mlk}}t 
-2\frac{d_{lk}}{AC_{lk}}z+t^2-m_k^2+i\rho \Big]}, 
\nonumber\\
\end{eqnarray}
where
\begin{eqnarray}
D_{mlk}   &=& 1-\frac{2\alpha_{lk}}{AC_{lk}} +
\frac{b_{lk}^2}{AC_{lk}^2}=-4\frac{(q_l-q_k)^2}{AC_{lk}^2} ,\nonumber\\
F_{nmlk} &=& \frac{C_{nlk}B_{mlk}-B_{nlk}C_{mlk}}
{A_{nlk}B_{mlk}-B_{nlk}A_{mlk}}.
\end{eqnarray}
We note that $D_{mlk}\in \mathbb{R}$ and $F_{nmlk}\in \mathbb{C}$.

With the definitions of $f_{kl}, f^-_{kl}$ in Eq.~(\ref{flk}) and
$g_{mlk},g^-_{mlk}$ in Eq.~(\ref{gmlk}), one again confirms that 
\begin{eqnarray} 
\label{Imgklm}
\mathrm{Im} \Big[D_{mlk}z^2-2\Big(\frac{A_{mlk}}{B_{mlk}}-\alpha_{lk}\Big)zt 
- 2\frac{C_{mlk}}{B_{mlk}}t -2\frac{d_{lk}}{AC_{lk}}z+t^2-m_k^2+i\rho \Big] > 0.  
\end{eqnarray}

We have just arrived at the two-fold integrations. In the next subsections, we will 
present the approach to calculate these integrals (\ref{yint1}, \ref{yint2}, \ref{yint3}, \ref{yint4})
in detail.
\subsection{The $t$-integration} 
To linearize $t$, we perform a shift  
\begin{equation}
\label{rotation}
\begin{aligned}
z = z' + \beta_{mlk} t' & & z' = \frac{z-\beta_{mlk}
t}{1-\beta_{mlk}\varphi_{mlk}}, \\
&\qquad \Longrightarrow \qquad& \\
t = t' + \varphi_{mlk} z' & & t' = \frac{t-\varphi_{mlk},
z}{1-\beta_{mlk}\varphi_{mlk}}.
\end{aligned}
\end{equation}
The Jacobian of this shift is
\begin{eqnarray}
 J=|1-\beta_{mlk}\varphi_{mlk}|.   \nonumber
\end{eqnarray}
To remove the quadratic term of $t$, we have to choose $\beta_{mlk}$ as the
roots of the equation
\begin{eqnarray}
 D_{mlk}\beta_{mlk}^{2}-2\Big(\frac{A_{mlk}}{B_{mlk}}-\alpha_{lk}\Big)\beta_{mlk}+1=0
\end{eqnarray}
or  these roots are written explicitly as
\begin{eqnarray}
\beta_{mlk}^{(1,2)}=\dfrac{\Big(\frac{A_{mlk}}{B_{mlk}}-\alpha_{lk}\Big) \pm 
\sqrt{\Big(\frac{A_{mlk}}{B_{mlk}}-\alpha_{lk}\Big)^2-D_{mlk}}}{D_{mlk}}.
\end{eqnarray} 
With this, the integrand written in terms of $t$ depends linearly on $t$ which is
\begin{equation}
\label{finallyz}
\mathcal{I}_{nmlk}(z, t) = \frac{1}{\left[z + \beta_{mlk} t + F_{nmlk}
\right]\left[ Q_{mlk} t + P_{mlk} t z + E_{mlk} z + Z_{mlk} z^2  
- m_k^2 + i \varrho \right]},
\end{equation}
with
\begin{eqnarray}
Q_{mlk}  & = & - 2 \left( \frac{C_{mlk}}{B_{mlk}}
+\frac{d_{lk}}{AC_{lk}}\beta_{mlk}\right), \\
P_{mlk} &=& -2 \left[\left(\frac{A_{mlk}}{B_{mlk}} - \alpha_{lk} \right) (1 +
\beta_{mlk}\varphi_{mlk}) - D_{mlk} \beta_{mlk} - \varphi_{mlk}  \right],
\\
E_{mlk} &=&  -2 \left( \frac{d_{lk}}{AC_{lk}} +
\frac{C_{mlk}}{B_{mlk}}\varphi_{mlk} \right),\\
Z_{mlk} &=&  D_{mlk} -2  \left(\frac{A_{mlk}}{B_{mlk}} - \alpha_{lk}
\right) \varphi_{mlk} + \varphi_{mlk}^2, \\
D_{mlk} &=& - 4 \frac{(q_l - q_k)^2}{AC^2_{lk}}, \\
F_{nmlk} &=&\frac{C_{nlk}B_{mlk} - B_{nlk} C_{mlk}}{A_{nlk} B_{mlk} - B_{nlk}
A_{mlk}}.
\end{eqnarray}

We will choose $\varphi_{mlk}$ as the root of the equation $Z_{mlk}=0$, 
\begin{eqnarray}
\varphi_{mlk}^{2}-2\Big(\frac{A_{mlk}}{B_{mlk}}-\alpha_{lk}\Big)\varphi_{mlk}+D_{mlk}=0
\end{eqnarray}
or their solutions are given by
\begin{eqnarray}
\varphi_{mlk}^{(1,2)}=\Big(\frac{A_{mlk}}{B_{mlk}} - \alpha_{lk}\Big) \pm 
\sqrt{\Big(\frac{A_{mlk}}{B_{mlk}}-\alpha_{lk}\Big)^2-D_{mlk}}. 
\end{eqnarray}
We note that the final result of $D_0$ is independent of the parameters $\varphi_{mlk}$ and $\beta_{mlk}$.
Without the loss of generality, we choose $\varphi_{mlk}=\varphi_{mlk}^{(1)}$ and
$\beta_{mlk}=\beta_{mlk}^{(1)}$
in the following calculation. In this case,  we have
\begin{eqnarray}
\label{pmlk}
 P_{mlk}=-4\Big[ \Big(\frac{A_{mlk}}{B_{mlk}}-\alpha_{lk}\Big)^2-D_{mlk}\Big]\beta_{mlk}.
\end{eqnarray}
The relations between $D_{mlk}$ with the external momenta are shown 
in Table~\ref{DP2}. 
\begin{table}[ht]
\begin{center}
\large
\begin{tabular}{|l|l|l|} \hline  
 $l=1,k=2$ &   $q_1-q_2=-p_2$        &        $D_{m12}=-4\frac{p_2^2}{AC_{lk}^2}$              \\ \hline
 $l=1,k=3$ &   $q_1-q_3=-p_2-p_3$ &        $D_{m13}=-4\frac{(p_2+p_3)^2}{AC_{lk}^2}$    \\ \hline
 $l=1,k=4$ &   $q_1-q_4=p_1$         &        $D_{m14}=-4\frac{p_1^2}{AC_{lk}^2}$               \\ \hline
 $l=2,k=3$ &   $q_2-q_3=-p_3$         &       $D_{m23}=-4\frac{p_3^2}{AC_{lk}^2}$               \\ \hline
 $l=2,k=4$ &   $q_2-q_4=p_1+p_2$ &        $D_{m24}=-4\frac{(p_1+p_2)^2}{AC_{lk}^2}$    \\ \hline
 $l=3,k=4$ &   $q_3-q_4=-p_4$         &        $D_{m34}=-4\frac{p_4^2}{AC_{lk}^2} $               \\ \hline
\end{tabular}
\end{center}
\caption{\label{DP2} $D_{mlk}$ are written in terms of external momenta.}
\label{tab1}
\end{table}

To follow the calculation easily, we would like to omit the index for the kinematic 
variables which appear in Eq.~(\ref{finallyz}) in the
remaining text of this paper.
\subsubsection{In the case of $D_{mlk} <0$}

\begin{figure}[ht]
 \begin{center}
\begin{pspicture}(-7, -2.5)(7, 2.5)
\psline{->}(-4,-2.5)(-4,2.5)
\psline{->}(4,-2.5)(4,2.5)
\pspolygon[fillstyle=crosshatch, fillcolor=gray,
linestyle=none, hatchcolor=lightgray](1.5,1.9)(4,0)(6.5,1.7)
\pspolygon[fillstyle=crosshatch,
linestyle=none, hatchcolor=lightgray, hatchangle=0, hatchwidth=1.2pt,
hatchsep=1.8pt](1.5,-1.7)(4, 0)(6.5,-1.9)
\pspolygon[fillstyle=hlines,
linestyle=none, linecolor=lightgray, hatchcolor=lightgray, hatchwidth=1.2pt,
hatchsep=1.8pt](6.5,-1.9)(4, 0)(6.5,1.7)
\pspolygon[fillstyle=hlines,
linestyle=none, linecolor=lightgray, hatchcolor=lightgray, hatchwidth=1.2pt,
hatchsep=1.8pt,hatchangle=135](1.5,-1.7)(4, 0)(1.5,1.9)
\psline{->}(-7,0)(-1,0)
\psline{->}(1,0)(7,0)
\psline(1.5,-1.7)(6.5,1.7)
\psline(1.5,1.9)(6.5,-1.9)
\rput(-3.8, 2.6){$t$}
\rput(4.2, 2.6){$t$}
\rput(-1.2,-0.2){$z$}
\rput(7,-0.2){$z$}
\rput(6.5, 2.2){$-\frac{1}{\beta_{mlk}z} $}
\rput(0.5, 1.8){$-\varphi_{mlk} z$}
\rput(6.,-0.5){$D^{++}$}
\rput(4.2,-1.2){$D^{+-}$}
\rput(4.2,1.2 ){$D^{-+}$}
\rput(2.,0.5)   {$D^{--}$}
\rput(-2.5, 1.5){$D^{++}$}
\rput(-2.5,-1.5){$D^{+-}$}
\rput(-5.2,1.5 ){$D^{-+}$}
\rput(-5.2, -1.5){$D^{--}$}
\psline{->}(-0.3,0)(0.3,0)
\end{pspicture}
\caption{\label{beta1} The integration region.}
\end{center}
\end{figure}
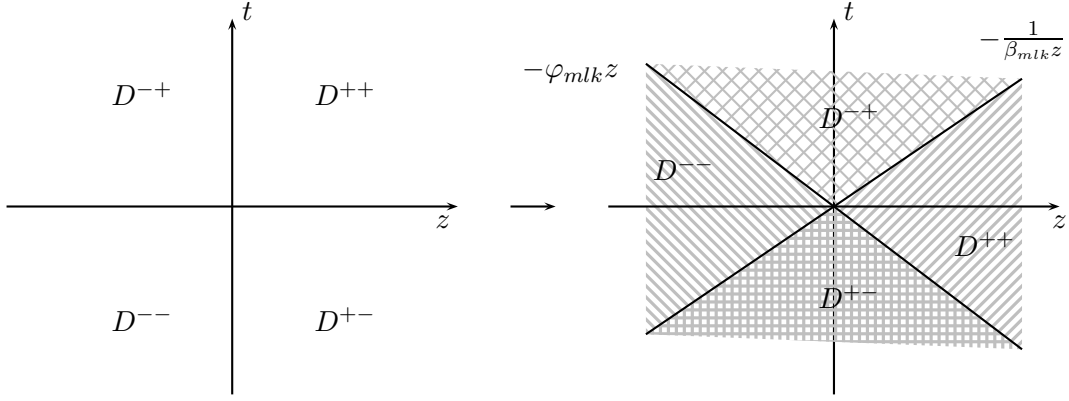

In this case, $\beta_{mlk} \leqslant 0$ and $\varphi_{mlk}\geqslant 0$. 
The integration region now looks as Fig.~\ref{beta1}. 

To integrate over $t$, one first splits the integrations written in terms of $t$ as follows
\begin{eqnarray}
D^{++} & \longrightarrow & \int\limits_{0}^{\infty} dz \int\limits_{-\varphi_{mlk}\; z}^{-1/\beta_{mlk}\; z}dt    
=  \int\limits_{0}^{\infty} dz \int\limits_{-\varphi_{mlk}\; z}^{\infty}dt
-\int\limits_{0}^{\infty}  dz \int\limits_{-1/\beta_{mlk}\; z}^{\infty} dt,  \nonumber\\
D^{+-} &\longrightarrow & \int\limits_{0}^{\infty} dz \int\limits^{-\varphi_{mlk}\; z}_{-\infty}dt
-\int\limits^{0}_{-\infty}  dz \int\limits^{-1/\beta_{mlk}\; z}_{-\infty} dt,   \nonumber\\
D^{-+} &\longrightarrow &\int\limits_{0}^{\infty} dz \int\limits^{\infty}_{-1/\beta_{mlk}\;z} dt
+\int\limits_{-\infty}^0 dz \int\limits^{\infty}_{-\varphi_{mlk}\; z} dt,  
\nonumber\\[1.5ex]
D^{--} &\longrightarrow & \int\limits_{-\infty}^0 dz \int\limits^{-\varphi_{mlk}\; z}_{-1/\beta_{mlk}\; z} dt 
=\int\limits_{-\infty}^0 dz \int\limits_{-1/\beta_{mlk}\; z}^\infty dt  
-  \int\limits_{-\infty}^0 dz\int\limits_{-\varphi_{mlk}\; z}^\infty dt.
\nonumber
\end{eqnarray}

We next rewrite the $t$-integrand in the form of
\begin{eqnarray}
 \mathcal{I}_{nmlk}(t, z) &=& \frac{1}{\beta (Q + P z)}\frac{1}{\left[\frac{  E z
-m_k^2 + i\varrho}{Q+ P z} - \frac{F + z}{\beta}\right]}
 \left\{  \frac{1}{t + \frac{F + z}{\beta}} - \frac{1}{t + \frac{ E z
-m_k^2 + i\varrho}{Q+ P z}} \right\}.
\end{eqnarray}

We are going to apply the formula (\ref{masterLog}) for calculating the $t$-integrations.
In order to use the formula (\ref{masterLog}), the integrand $\mathcal{I}_{nmlk}(t, z) $ 
must have no poles in real $t$-axes. However, in the case of real masses, $F_{nmlk}$ are real, and so 
$\mathcal{I}_{nmlk}(t, z)$ may have poles in negative real $t$-axes. To treat this problem, we make 
$F_{nmlk}\rightarrow F_{nmlk}+i\rho'$ with $\rho'\geqslant 0$. The final result is obtained by taking
$\rho'\rightarrow 0$.  Using the master integral (\ref{masterLog}), one gets
\begin{eqnarray}
\label{i3}
\int\limits_{-\infty}^{\sigma z} dt \; \mathcal{I}_{nmlk}(t,z) &=&
\frac{1}{\beta (Q + P
z)}\frac{1}{\left[\frac{ E z -m_k^2 + i\varrho}{Q+ P z} - \frac{F +
z}{\beta}\right]}\left\{ \ln\left( - \frac{(1 + \beta\sigma) z +F  }{\beta}
\right) \right. \nonumber   \\[1ex]
&& \hspace{0cm} \left. - \ln\left( - \frac{  P \sigma z^2 + (E + Q\sigma) z
-m_k^2 + i\varrho }{Q+ P z} \right)
\right\},
\end{eqnarray}
and 
\begin{eqnarray}
\label{i4}
\int\limits_{\sigma z}^{\infty} dt\; \mathcal{I}_{nmlk}(t, z) &=& \frac{1}{\beta (Q + P
z)}\frac{1}{\left[\frac{ E z -m_k^2 + i\varrho}{Q+ P z} - \frac{F +
z}{\beta}\right]}\left\{- \ln\left( \frac{(1 + \beta\sigma) z +F  }{\beta}
\right) \right. \nonumber   \\[1ex]
&& \hspace{0cm} \left. + \ln\left( \frac{ P\sigma z^2 + (E + Q\sigma) z -m_k^2
+ i\varrho }{Q+ P z} \right)
\right\}. 
\end{eqnarray}
Where $\sigma$ will be $-\varphi_{mlk}$ or $-\frac{1}{\beta_{mlk}}$.

With the help of (\ref{i3}, \ref{i4}), one obtains
\begin{eqnarray}
  \dfrac{D^{++}}{i\pi^2}  &=& \bigoplus_{nmlk} f_{lk}g_{mlk}\int\limits_{0}^{\infty} dz \left\{ \int\limits_{-\varphi z}^\infty dt -
 \int\limits_{-1/\beta z}^\infty dt \right\}\mathcal{I}_{nmlk}(t,z) \\
 &=&  \bigoplus_{nmlk} f_{lk}g_{mlk}  \int\limits_{0}^{\infty} dz\; G(z) \left\{-\ln\left(\frac{(1-\beta\varphi)z
 +F}{\beta}    \right) +  \ln\left(\frac{F}{\beta} \right)
 \right.\nonumber\\
 &&\hspace{4.5cm} + \ln\left(\frac{-P
 \varphi z^2 +(E - Q \varphi) z -m_k^2 + i\varrho }{Q + P z} \right)\nonumber\\
 && \hspace{4.5cm}\left.  - \ln\left( \frac{ -\frac{P}{\beta} z^2 + (E -
 \frac{Q}{\beta}) z -m_k^2 + i\varrho }{Q + P z} \right)  \right\}; \nonumber
 \end{eqnarray}
 \begin{eqnarray}
\dfrac{D^{+-}}{i\pi^2} &=& -\bigoplus_{nmlk} f_{lk}g^-_{mlk} \left\{\int \limits_{0}^{\infty} dz
\int\limits^{-\varphi z}_{-\infty} dt  
+  \int\limits^{0}_{-\infty} dz \int^{-1/\beta z}_{-\infty}dt \right\} \mathcal{I}_{nmlk}(t,z)\nonumber\\
& =&  - \bigoplus_{nmlk} f_{lk}g^-_{mlk} \left\{ \int\limits_{0}^{\infty} dz\; G(z) 
\left[\ln\left(-\frac{(1-\beta\varphi)z +F}{\beta}    \right)\right. \right. \\
&   & \hspace{4.5cm} \left. - \ln\left(-\;\frac{-P
\varphi z^2 + (E - Q \varphi) z -m_k^2 + i\varrho }{Q + P z} \right)    \right]
+ \nonumber\\
& &\hspace{1cm}\left. \int^{0}_{-\infty} dz \; G(z)\left[\ln\left(-\frac{F}{\beta} \right) - \ln\left(-
 \;\frac{-\frac{P}{\beta} z^2 + (E - \frac{Q}{\beta}) z -m_k^2 + i\varrho }{Q + P z} \right) \right] 
 \right\}; \nonumber
\end{eqnarray}
\begin{eqnarray}
 \dfrac{D^{-+}}{i\pi^2}  &=& -\bigoplus_{nmlk} f^-_{lk}g_{mlk} \left\{ \int\limits_{0}^{\infty} dz  
 \int\limits_{-1/\beta z}^{\infty}dt
+   \int\limits^{0}_{-\infty} dz \int_{-\varphi z}^{\infty} dt  \right\} \mathcal{I}_{nmlk}(t,z)
\nonumber\\
&=&  -\bigoplus_{nmlk} f^-_{lk}g_{mlk} \left\{  \int\limits_{0}^{\infty} dz\;
G(z)\left[-\ln\left(\frac{F}{\beta} \right) +   \ln\left(\frac{
-\frac{P}{\beta} z^2 + (E - \frac{Q}{\beta}) z -m_k^2 + i\varrho }{Q + P z}
\right) \right] + \right.\nonumber\\
&&\left. \int\limits^{0}_{-\infty} dz \; G(z)\left[- \ln\left( \frac{(1-\beta\varphi)z
+F}{\beta}    \right) + \ln\left(  \frac{-P
\varphi z^2 +(E - Q \varphi) z -m_k^2 + i\varrho
}{Q + P z} \right)    \right]  \right\}; \nonumber\\
\end{eqnarray}
\begin{eqnarray}
\dfrac{D^{--}}{i\pi^2}  &=& \bigoplus_{nmlk} f^-_{lk}g^-_{mlk} \int\limits^{0}_{-\infty} dz \left\{ \int\limits_{-1/\beta
z}^\infty dt  - \int\limits_{-\varphi z}^\infty dt \right\} \mathcal{I}_{nmlk}(t,z) \nonumber\\
&=& \bigoplus_{nmlk} f^-_{lk}g^-_{mlk} \int\limits^{0}_{-\infty} dz \; G(z) \left\{- \ln\left(
\frac{F}{\beta} \right) +   \ln\left(\frac{-\frac{P}{\beta} z^2 + (E -
\frac{Q}{\beta}) z -m_k^2 + i\varrho }{Q + P z} \right)  \right.\nonumber\\
&&\hspace{2cm} \left. + \ln\left(\frac{(1-\beta\varphi)z +F}{\beta}    \right)
-   \ln\left(\frac{ -P\varphi z^2 +  (E - Q \varphi) z -m_k^2 + i\varrho
}{Q + P z} \right) \right\}.
\nonumber\\
\end{eqnarray}
Here the $G(z)$ function is given
\begin{eqnarray}
\label{GZ}
 G(z)=\frac{1}{\beta(Ez-m_k^2+i\rho)-(Q+Pz)(F+z)}.
\end{eqnarray}
With the help of $i\rho'$, all the logarithmic functions 
which appear in the $z$-integrands now are well-defined in 
$z$-complex plane. Summing up the above terms, one obtains
 \begin{eqnarray}
 \label{zint-ln}
\dfrac{D_0}{i\pi^2}  &=& \sum_{k=1}^{4}\sum_{\substack{l=1\\k\neq l}}^{4}
\sum_{\substack{m=1\\m\neq l\\m\neq k}}^{4}\frac{\Big(1-\delta_{lk}(AC_{lk})\Big)\Big(1-\delta_{lk}(B_{mlk})\Big)}
{AC_{lk}(B_{mlk}A_{nlk}-B_{nlk}A_{mlk})} \left|1 - \beta_{mlk} \varphi_{mlk} \right| \times\nonumber\\ 
&&\times \left[ \hspace{0.4cm}\int\limits_{0}^{\infty} dz \;
G(z) \left\{ (f_{lk}g_{mlk} + f^-_{lk}g_{mlk} )\ln\left(\frac{F}{\beta} \right)
\right.\right.\nonumber\\
&& - f_{lk}g_{mlk}  \ln\left(\frac{(1-\beta\varphi)z +F}{\beta}\right) -
f_{lk}g^-_{mlk}  \ln\left(-\;\frac{(1-\beta\varphi)z
+F}{\beta}\right)\nonumber\\
&& - (f_{lk}g_{mlk} + f^-_{lk}g_{mlk} ) \ln\left(\frac{ -\frac{P}{\beta}
z^2 + (E - \frac{Q}{\beta}) z -m_k^2 + i\varrho }{Q + P z} \right)\nonumber\\
&& +  f_{lk}g_{mlk} \ln\left(\frac{-P \varphi z^2 + (E - Q \varphi) z -m_k^2 +
i\varrho }{Q + P z} \right)\nonumber\\
&&\left.  + f_{lk}g^-_{mlk} \ln\left(-\;\frac{-P \varphi z^2 + (E - Q \varphi)
z -m_k^2 + i\varrho }{Q + P z} \right)
\right\}\nonumber\\[2ex]\nonumber
&& \hspace{1cm} + \int_{-\infty}^0 dz \; G(z) \left\{- f^-_{lk}g^-_{mlk}
\ln\left( \frac{F}{\beta} \right) - f_{lk}g^-_{mlk} \ln\left(- \frac{F}{\beta}
\right)\right. \\
&& + (f^-_{lk}g^-_{mlk} + f^-_{lk}g_{mlk} ) \ln\left(\frac{(1-\beta\varphi)z
+F}{\beta}\right) \nonumber\\
&& + f^-_{lk}g^-_{mlk}\ln\left( \frac{ -\frac{P}{\beta} z^2 + (E -
\frac{Q}{\beta}) z -m_k^2 + i\varrho }{Q + P z} \right)                                          \\
&& + f_{lk}g^-_{mlk}\ln\left(-\frac{-\frac{P}{\beta} z^2 + (E -
\frac{Q}{\beta}) z -m_k^2 + i\varrho }{Q + P z} \right) \nonumber\\
&&\left. \left. - (f^-_{lk}g^-_{mlk} + f^-_{lk}g_{mlk} )  \ln\left(\frac{- P
\varphi z^2 + (E - Q \varphi) z -m_k^2 + i\varrho }{Q + P z} \right)\right\}
\;\hspace{0.4cm}  \right].
\nonumber
\end{eqnarray}

In the next steps, we are now going to calculate the $z$-integrals. We realize that
\begin{eqnarray}
\label{Im(e-q)}
\mathrm{Im}\Big(E_{mlk}-Q_{mlk}\varphi_{mlk}\Big)=
\begin{cases}
        \geq 0,   &\text{with}    \;\;  ++,+-;   \\\\
        \leq 0    &\text{with}    \;\;  -+,--;
\end{cases}
\end{eqnarray}
and
\begin{eqnarray}
\label{Im(e-qb)}
\mathrm{Im}
\Big(E_{mlk}-\frac{Q_{mlk}}{\beta_{mlk}}\Big)=
\begin{cases}
        \geq 0,   &\text{with}    \;\;  ++,+-; \\\\
        \leq  0    &\text{with}     \;\;  -+,--. 
\end{cases}
\end{eqnarray}
Where we use the notations $++,+-, \cdots, --$ which are corresponding to 
the appearance of $f_{lk}g_{mlk},f_{lk}g_{mlk}^-, \cdots, f_{lk}^-g_{mlk}^-$ 
in the mentioned formulae.

From Eqs.~(\ref{Im(e-q)}, \ref{Im(e-qb)}), we also confirm that
\begin{eqnarray}
\label{Im}
\text{Im}\Big(P\sigma z^2+(E+Q\sigma)z-m_k^2+i\rho\Big)\geq0
\end{eqnarray}
with  $\sigma=-\varphi_{mlk}$ or  $\sigma=-1/\beta_{mlk}$.

From the formula~(\ref{pmlk}), we realize that $P_{mlk}$ is a nonzero real in 
the case $D_{mlk}<0.$ We rewrite $ G(z)$ as follows
\begin{eqnarray}
G(z)&=& \frac{1}{-P(z-T_1)(z-T_2)},
\end{eqnarray}
with
\begin{eqnarray}
T_1&=&\dfrac{(Q+PF-\beta E)}{-2P}+\dfrac{\sqrt{(Q+PF-\beta E)^2-4P(QF+\beta m_k^2-i\beta\rho)}}{-2P}, 
\\
T_2&=&\dfrac{(Q+PF-\beta E)}{-2P}-\dfrac{\sqrt{(Q+PF-\beta E)^2-4P(QF+\beta m_k^2-i\beta\rho)}}{-2P}.
\end{eqnarray}
Defining the arguments of logarithmic functions are
\begin{eqnarray}
\label{SZ}
 S(\sigma,z)&=&P\sigma z^2+(E+Q\sigma)z-m_k^2+i\rho \nonumber\\
&=&P\sigma (z-Z_{1\sigma})(z-Z_{2\sigma}), 
\end{eqnarray}
with
\begin{eqnarray}
\label{zphibeta}
Z_{1\varphi}&=& \dfrac{(E-Q\varphi)+\sqrt{(E-Q\varphi)^2-4P\varphi(m_k^2-i\rho)}}{2P\varphi}, \\
Z_{2\varphi}&=& \dfrac{(E-Q\varphi)-\sqrt{(E-Q\varphi)^2-4P\varphi(m_k^2-i\rho)}}{2P\varphi},  \\
Z_{1\beta}   &=& \dfrac{(E-\frac{Q}{\beta})+\sqrt{(E-\frac{Q}{\beta})^2-4\frac{P}{\beta}(m_k^2-i\rho)}}{\frac{2P}{\beta}},\\
Z_{2\beta}   &=& \dfrac{(E-\frac{Q}{\beta})-\sqrt{(E-\frac{Q}{\beta})^2-4\frac{P}{\beta}(m_k^2-i\rho)}}{\frac{2P}{\beta}}.
\end{eqnarray}

In order to perform the $z$-integrals, we decompose logarithmic functions 
in the $z$-integrands, as Eq.~(\ref{logdecompose}).  In special, one has 
\begin{eqnarray}
\label{decompose}
\ln\Big(\frac{ S(\sigma,z)}{Pz+Q}\Big)&=&\ln(P\sigma z-P\sigma Z_{1\sigma})
+ \ln(z-Z_{2\sigma})-\ln(Pz+Q)\\
&&+2\pi i\theta[\text{Im}(P\sigma Z_{1\sigma})]\theta[\text{Im}( Z_{2\sigma})] - 
2\pi i \theta[-\text{Im}(Q)]\theta\Big[-\text{Im} \left( \frac{S(\sigma,z)}{Pz+Q} \right) \Big], 
\nonumber
\end{eqnarray}
and
\begin{eqnarray}
\label{decomposelog1}
\ln\Big(\frac{ -S(\sigma,z)}{Pz+Q}\Big)&=&\ln(-P\sigma z+P\sigma Z_{1\sigma})+\ln(z-Z_{2\sigma})-\ln(Pz+Q)\\
&&-2\pi i\theta[\text{Im}(P\sigma Z_{1\sigma})]\theta[-\text{Im}( Z_{2\sigma})]+
2\pi i \theta[\text{Im}(Q)]\theta\Big[-\text{Im} \left( \frac{S(\sigma,z)}{Pz+Q} \right)\Big]. \nonumber
\end{eqnarray}

By determining that $\text{Im}\left(\pm\dfrac{S(\sigma,z)}{Pz+Q}\right)$  is independent  of $\sigma$ and using
the formulae~(\ref{decompose}, \ref{decomposelog1}), $D_0$ can be presented in the form 
\begin{eqnarray}
\label{zlntdecompose2}
\dfrac{D_0}{i\pi^2} &=&\bigoplus_{nmlk} 
 \int\limits_{0}^{\infty} dz G(z)\Bigg\{\;\; \Omega^{+}_{nmlk} -f_{lk}g_{mlk}\; \ln\Big(\frac{1-\beta\varphi}{\beta}z+\frac{F}{\beta}\Big)\nonumber\\
&&\hspace{0.0cm}-f_{lk}g^{-}_{mlk}\;\ln\Big(\frac{-(1-\beta\varphi)}{\beta}z-\frac{F}{\beta}\Big)-(f_{lk}g_{mlk}+f^{-}_{lk}g_{mlk})
\ln\Big(\frac{-Pz}{\beta}+\frac{P Z_{1\beta}}{\beta}\Big)\nonumber\\
&&\hspace{0.0cm}-(f_{lk}g_{mlk}+f^{-}_{lk}g_{mlk})\ln\Big(z-Z_{2\beta}\Big)+f_{lk}g_{mlk}\; \ln( -P\varphi z+P\varphi Z_{1\varphi})\nonumber\\
&&\hspace{0.0cm}+f_{lk}g_{mlk}\; \ln(  z- Z_{2\varphi})+f_{lk}g^-_{mlk}\; \ln(P\varphi z-P\varphi Z_{1\varphi})\nonumber\\
&&\hspace{0.0cm}+f_{lk}g^-_{mlk}\; \ln(  z- Z_{2\varphi}) +(f^-_{lk}g_{mlk}-f_{lk}g^-_{mlk})\;\ln(Pz+Q)\hspace{0.5cm}\Bigg\}\nonumber\\
&&+\bigoplus_{nmlk}\int\limits^{0}_{-\infty} dz G(z)\Bigg\{ \;\; \Omega^{-}_{nmk}+f^-_{lk}g^-_{mlk}\; \ln\Big(z-Z_{2\beta}\Big)               
\\
&&\hspace{0.0cm}+f_{lk}g^-_{mlk}\; \ln\Big( \frac{Pz}{\beta}-\frac{P Z_{1\beta}}{\beta}\Big)+(f^-_{lk}g^-_{mlk} +f^-_{lk}g_{mlk})\;
\ln\Big(\frac{1-\beta\varphi}{\beta}z+\frac{F}{\beta}\Big)\nonumber\\
&&\hspace{0.0cm}+f^-_{lk}g^-_{mlk}\; \ln\Big( \frac{-Pz}{\beta}+\frac{P Z_{1\beta}}{\beta}\Big)-(f^-_{lk}g^-_{mlk}+f^-_{lk}g_{mlk})\;
\ln(-P\varphi z+P\varphi Z_{1\varphi})\nonumber\\
&&\hspace{0.0cm}+f^-g^-\; \ln\Big(z-Z_{2\beta}\Big)+(f^-_{lk}g_{mlk}-f_{lk}g^-_{mlk})\; \ln(Pz+Q)\nonumber\\
&&-(f^-_{lk}g^-_{mlk}+f^-_{lk}g_{mlk})\; \ln(z-Z_{2\varphi})\hspace{0.7cm}\Bigg\} \nonumber\\
&&+ 2\pi\; i\;  \bigoplus_{nmlk} \Big(f_{lk}g^-_{mlk} \theta[\text{Im}(Q)]+f^-_{lk}g_{mlk}\theta[-\text{Im}(Q)]\Big)
\int\limits^{\infty}_{-\infty} dz\; G(z)\theta \left[-\text{Im}\left(\dfrac{S(\sigma,z)}{Pz+Q}\right)\right]. \nonumber
\end{eqnarray}
We first emphasize that $\ln(Pz+Q) $ might have poles in negative real-axes in the real 
mass cases. However, in these cases
 $f_{lk}=f_{lk}^-=1$ and $g_{mlk}=g_{mlk}^-=1$. As a result, one checks that
 $(f^{-}_{lk}g_{mlk}^{-} -f_{lk}g_{mlk}) \ln( Pz+Q)=0$.  Thus
 we don't need to make $Q\longrightarrow Q+i\rho'$ as $F_{nmlk}$ case.  
 Secondly, the $z$-integrals now are splitted into three basic integrals which are
\begin{eqnarray}
 \int\limits_{0}^{\infty} \;G(z)dz ; \; \; \; \int\limits_{0}^{\infty} \;\ln(az + b)G(z)dz ; \; \; \; \int\limits_{-\infty}^{\infty} 
 \theta\Big[-\text{Im} \left(\dfrac{S(\sigma,z)}{Pz+Q}\right)\Big] G(z)dz.\nonumber
\end{eqnarray}
These integrals can be calculated in concrete in the Appendix.
%
\subsubsection{In the case of $0<D_{mlk}\leqslant  
\Big(\frac{A_{mlk}}{B_{mlk}}-\alpha_{lk}\Big)^2$ 
and $\frac{A_{mlk}}{B_{mlk}}-\alpha_{lk} < 0$}
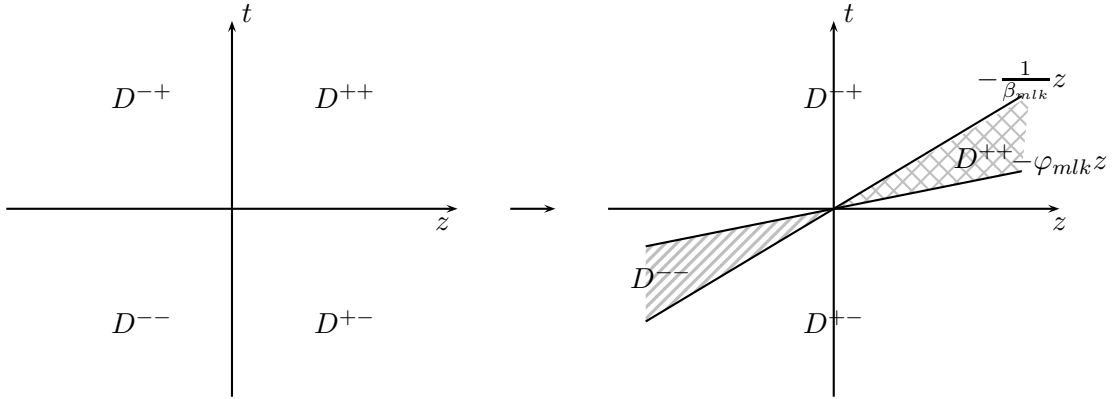
\begin{figure}[ht]
	 \begin{center}
	\begin{pspicture}(-7, -2.5)(7, 2.5)
	\psline{->}(-4,-2.5)(-4,2.5)
	\psline{->}(4,-2.5)(4,2.5)
	\psline{->}(-7,0)(-1,0)
	\psline{->}(1,0)(7,0)
	\pspolygon[fillstyle=crosshatch, fillcolor=gray,
	linestyle=none, hatchcolor=lightgray](6.6,1.5)(4,0)(6.5,0.5)
	\pspolygon[fillstyle=hlines,linestyle=none, linecolor=lightgray, hatchcolor=lightgray, hatchwidth=1.2pt,
 	hatchsep=1.8pt](1.5,-0.5)(4, 0)(1.5,-1.5)
	\psline(1.5,-0.5)(6.5,0.5)
	\psline(1.5,-1.5)(6.5,1.5)
	\rput(-3.8, 2.6){$t$}
	\rput(4.2, 2.6){$t$}
	\rput(-1.2,-0.2){$z$}
	\rput(7,-0.2){$z$}
	\rput(7., 0.6){$-\varphi_{mlk}z$}
 	\rput(6.5, 1.7){$-\frac{1}{\beta_{mlk}}z$}
	\rput(6.,0.7){$D^{++}$}
	\rput(1.7,-0.9){$D^{--}$}
 	\rput(4.0,-1.5 ){$D^{+-}$}
 	\rput(4.,1.5){$D^{-+}$}	
	\rput(-2.5, 1.5){$D^{++}$}
	\rput(-2.5,-1.5){$D^{+-}$}
	\rput(-5.2,1.5 ){$D^{-+}$}
	\rput(-5.2, -1.5){$D^{--}$}
	\psline{->}(-0.3,0)(0.3,0)
	\end{pspicture}
	\end{center}
	 \caption{\label{casefig3} The integration region.}
	\end{figure}

	In this case, we have
	\begin{eqnarray}
	 \beta_{mlk}&=&\dfrac{\Big(\frac{A_{mlk}}{B_{mlk}}-\alpha_{lk}\Big)+ \sqrt{\Big(\frac{A_{mlk}}{B_{mlk}}-\alpha_{lk}\Big)^2-D_{mlk}}}{D_{mlk}} < 0,
	 \\
	 \varphi_{mlk}&=&\Big(\dfrac{A_{mlk}}{B_{mlk}}-\alpha_{lk}\Big)+ \sqrt{\Big(\frac{A_{mlk}}{B_{mlk}}-\alpha_{lk}\Big)^2-D_{mlk}} <0.	
	\end{eqnarray}
	It is easy to check that
	 $-\varphi_{mlk}-(-\frac{1}{\beta_{mlk}}) = -\sqrt{\Big(\frac{A_{mlk}}{B_{mlk}}-\alpha_{lk}\Big)^2-D_{mlk}}\leqslant0$. 
	 Therefore, the integration region now looks like Fig.~\ref{casefig3}.
In this case, the integration region of $D_0$ is similar to $D_{mlk}<0$ case. 
As a result, the analytical calculation of $D_0$ in this case is same to the case of $D_{mlk}<0$.\\
\subsubsection{In the case of $0<D_{mlk}\leqslant          
\Big(\frac{A_{mlk}}{B_{mlk}}-\alpha_{lk}\Big)^2$ and        
$\frac{A_{mlk}}{B_{mlk}}-\alpha_{lk}> 0$}                   
\begin{figure}[ht]
	\begin{center}
	\begin{pspicture}(-7, -2.5)(7, 2.5)
	\psline{->}(-4,-2.5)(-4,2.5)
	\psline{->}(4,-2.5)(4,2.5)
	\psline{->}(-7,0)(-1,0)
	\psline{->}(1,0)(7,0)
	\pspolygon[fillstyle=crosshatch, fillcolor=gray,
	linestyle=none, hatchcolor=lightgray](1.4,1.5)(4,0)(1.2,0.5)
	\pspolygon[fillstyle=hlines,linestyle=none, linecolor=lightgray, hatchcolor=lightgray, hatchwidth=1.2pt,
 	hatchsep=1.8pt](6.5,-1.5)(4, 0)(6.5,-0.5)
	\psline(1.5,0.5)(6.5,-0.5)
	\psline(1.5,1.5)(6.5,-1.5)
	\rput(-3.8, 2.6){$t$}
	\rput(4.2, 2.6){$t$}
	\rput(-1.2,-0.2){$z$}
	\rput(7,-0.2){$z$}
	\rput(0.8, 0.5){$-\frac{1}{\beta_{mlk}}z$}
 	\rput(1.5, 1.7){$-\varphi_{mlk}z$}
	\rput(4.,1.){$D^{++}$}
	\rput(4.,-1.2){$D^{--}$}
 	\rput(6.2,-0.8 ){$D^{-+}$}
 	\rput(2.,0.8){$D^{+-}$}
	\rput(-2.5, 1.5){$D^{++}$}
	\rput(-2.5,-1.5){$D^{+-}$}
	\rput(-5.2,1.5 ){$D^{-+}$}
	\rput(-5.2, -1.5){$D^{--}$}
	\psline{->}(-0.3,0)(0.3,0)
	\end{pspicture}
	\end{center}
	 \caption{\label{casefig4} The integration region.}
	\end{figure}
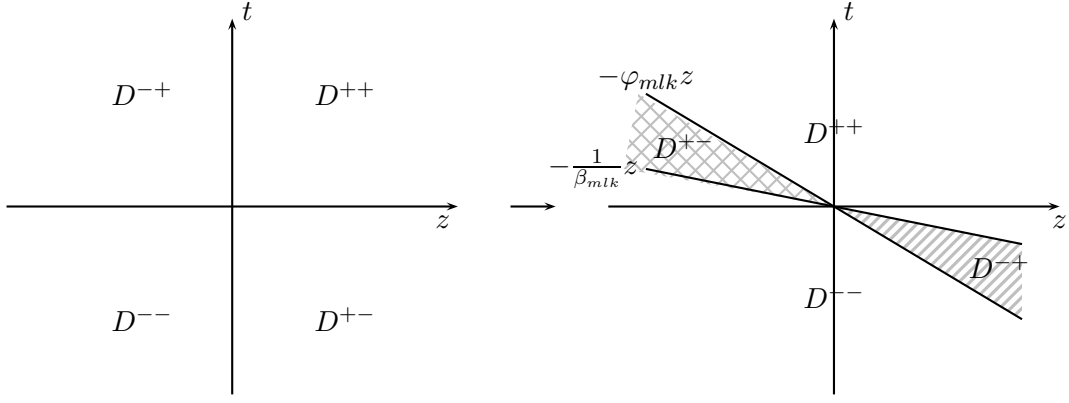

In this case, $\beta_{mlk}$ and $\varphi_{mlk}$ are positive and we confirm that 
\begin{eqnarray}
(-\varphi_{mlk})-(-\frac{1}{\beta_{mlk} } )=-\sqrt{\Big(\frac{A_{mlk}}{B_{mlk}}-\alpha_{lk}\Big)^2-D_{mlk}}\leqslant 0. 
\end{eqnarray}
Therefore, the integration region now looks like Fig.~\ref{casefig4}.

Applying the same procedure, $D_0$ reads
\begin{eqnarray}
\dfrac{D_0}{i\pi^2}&=&\bigoplus_{nmlk} 
\int\limits_{0}^{\infty}\;dz\;G(z)\Bigg\{ \;\; \Theta_{nmlk}+ f^{-}_{lk}g_{mlk}\ln\Big( \frac{1-\beta\varphi}{\beta}z +\frac{F}{\beta}\Big)\nonumber\\
&&\hspace{0.5cm}+f^{-}_{lk}g_{mlk}^{-}\ln\Big(-\frac{1-\beta\varphi}{\beta}z-\frac{F}{\beta}\Big) -f^{-}_{lk}g_{mlk}\ln\Big(-P\varphi z + P\varphi Z_{1\sigma}\Big)\nonumber\\
&&\hspace{0.5cm}-f^{-}_{lk}g^{-}_{mlk}\ln\Big(P\varphi z - P\varphi Z_{1\sigma}\Big)+(f_{lk}g_{mlk}+f^{-}_{lk}g_{mlk})\; \ln\Big( -\frac{P}{\beta}z + \frac{P}{\beta} Z_{1\beta}\Big)\nonumber\\
&&\hspace{0.5cm}-(f^{-}_{lk}g_{mlk}+f^{-}_{lk}g^{-}_{mlk}) \ln(z- Z_{2\varphi})+(f_{lk}g_{mlk}+f_{lk}^{-}g_{mlk})\ln(z- Z_{2\beta})\nonumber\\
&&\hspace{0.5cm}+(f_{lk}^{-}g_{mlk}^{-}-f_{lk}g_{mlk} )\ln(Pz+Q)\hspace{0.5cm}\Bigg\}                                                                            \\
&&+\bigoplus_{nmlk}\int\limits_{-\infty}^{0}\;dz\;G(z)\Bigg\{ \;\;\Theta_{nmlk}^{-} - (f_{lk}g_{mlk}^{-}+ f_{lk}g_{mlk})\ln \Big(\frac{(1-\beta\varphi)}{\beta}z + \frac{F}{\beta}\Big)\nonumber\\
&&\hspace{0.5cm}-f_{lk}g^{-}_{mlk}\ln \Big(-\frac{P}{\beta}z + \frac{P}{\beta} Z_{1\beta}\Big)-f_{lk}^{-}g^{-}_{mlk}\ln \Big(\frac{P}{\beta}z - \frac{P}{\beta} Z_{1\beta}\Big) \nonumber\\
&&\hspace{0.5cm}+ (f_{lk}g_{mlk} + f_{lk}g_{mlk}^{-})\ln(-P\varphi z+ P\varphi Z_{1\varphi})-(f_{lk}g_{mlk}^{-} + f_{lk}^{-}g_{mlk}^{-})\ln(z-Z_{2\beta})\nonumber\\
&&\hspace{0.5cm}+(f_{lk}g_{mlk} + f_{lk}g_{mlk}^{-} )\ln(z- Z_{2\varphi}) +(f^{-}_{lk}g_{mlk}^{-} -f_{lk}g_{mlk}) \ln( Pz+Q)\hspace{0.5cm}\Bigg\}\nonumber\\
&&-2\pi i\;\bigoplus_{nmlk} \Big(f_{lk}g_{mlk} \theta[-\text{Im}(Q)] + f^{-}_{lk}g^{-}_{mlk} \theta[\text{Im}(Q)]\Big)
\int\limits_{-\infty}^{\infty} \theta\left[ - \text{Im}\left( \dfrac{S(\sigma, z)}{P z+Q}\right)\right]
\;G(z) dz.
\nonumber 
\end{eqnarray}
Where the new kinematic variables introduced in this formula 
are
\begin{eqnarray}
 \Theta_{nmlk}&=& -(f_{lk}g_{mlk} + f_{lk}^{-}g_{mlk}) \ln\Big(\frac{F}{\beta}\Big) 
 - 2\pi i f_{lk}^{-}g_{mlk}\theta[-\text{Im}(P\varphi Z_{1\varphi})]\theta[\text{Im}(Z_{2\varphi})]\nonumber\\
&&+ 2\pi i f^{-}_{lk} g_{mlk}^{-}\theta[-\text{Im}(P\varphi Z_{1\varphi})]\theta[-\text{Im}(Z_{2\varphi})]\nonumber\\
&& +2\pi i (f_{lk}g_{mlk}+f^{-}_{lk}g_{mlk})\theta\Big[ \text{Im}(-\frac{PZ_{1\beta}}{\beta})\Big]\theta[\text{Im}(Z_{2\beta})],  
\end{eqnarray}
and
\begin{eqnarray}
 \Theta_{nmlk}^{-}&=& f_{lk}g_{mlk}^{-}\ln\Big(\frac{F}{\beta}\Big)+ f^{-}_{lk}g^{-}_{mlk}\ln\Big(-\frac{F}{\beta}\Big)
 -2\pi i f_{lk}g^{-}_{mlk} \theta\Big[-\text{Im}\left( \frac{PZ_{1\beta}}{\beta}\right)\Big]\theta[\text{Im}(Z_{1\beta})]\nonumber\\
&&+2\pi i f_{lk}^{-}g^{-}_{mlk} \theta\Big[-\text{Im} \left(\frac{PZ_{1\beta}}{\beta} \right) \Big]\theta[-\text{Im}(Z_{1\beta})]\nonumber\\
&&+2\pi i \Big(f_{lk}g_{mlk}+ f_{lk}g^{-}_{mlk} \Big) \theta\Big[-\text{Im}(P\varphi Z_{1\varphi})\Big]\theta[\text{Im}(Z_{2\varphi})].
\end{eqnarray}
The last integrals written in terms of $z$ will be evaluated
by means of the basic integrals which are presented in Appendix.

\subsubsection{In the case of $D_{mlk} > \small \Big(\frac{A_{mlk}}{B_{mlk}}-\alpha_{lk}\Big)^2$} 
From Eq.~(\ref{zint-ln}), each term relating to the
following integral will be presented as
\begin{eqnarray}
 \mathcal{T}_{12}^+ &=& C_{\sigma}\int\limits_{0}^{\infty} \dfrac{ \ln\left(\frac{-P \sigma z^2 + (E - Q \sigma) z -m_k^2 +
i\varrho }{Q + P z} \right) }{(z-T_1)(z-T_2)}dz \nonumber\\
&=& \dfrac{{C_{\sigma}}}{T_1-T_2}\int\limits_{0}^{\infty} \left\{
\dfrac{ \ln\left(\frac{-P \sigma z^2 + (E - Q \sigma) z -m_k^2 +
i\varrho }{Q + P z} \right) }{z-T_1}dz - \dfrac{ \ln\left(\frac{-P \sigma z^2 + (E - Q \sigma) z -m_k^2 +
i\varrho }{Q + P z} \right) }{z-T_2}dz \right\}\nonumber\\
&=& \dfrac{{C_{\sigma}}}{T_1-T_2}\int\limits_{0}^{\infty} \left\{
\dfrac{ \ln\left(\frac{-P \sigma z^2 + (E - Q \sigma) z -m_k^2 +
i\varrho }{Q + P z} \right) }{z-T_1}dz  - \dfrac{ \ln\left(\frac{-P \sigma T_1^2 + (E - Q \sigma) T_1 -m_k^2 +
i\varrho }{Q + P T_1} \right) }{z-T_1}dz \right\}\nonumber\\
&&+ \dfrac{{C_{\sigma}}}{T_2-T_1}\int\limits_{0}^{\infty} \left\{
\dfrac{ \ln\left(\frac{-P \sigma z^2 + (E - Q \sigma) z -m_k^2 +
i\varrho }{Q + P z} \right) }{z-T_2}dz  - \dfrac{ \ln\left(\frac{-P \sigma T_2^2 + (E - Q \sigma) T_2 -m_k^2 +
i\varrho }{Q + P T_2} \right) }{z-T_2}dz \right\}\nonumber\\
&&+\dfrac{{C_{\sigma}}}{T_1-T_2}\sum\limits_{i=1}^2\;
\int\limits_{0}^{\infty} \dfrac{ \ln\left(\frac{-P \sigma T_i^2 + (E - Q \sigma) T_i -m_k^2 +
i\varrho }{Q + P\;T_i} \right) }{z-T_i}dz. 
\end{eqnarray}
Where $C_{\sigma}$ are coefficients in front of the mentioned integrals. We also apply the 
same trick for each term relates to 
\begin{eqnarray}
  \mathcal{T}_{12}^- &=& C_{\sigma}\int\limits_{-\infty}^{0} \dfrac{ \ln\left(\frac{-P \sigma z^2 + (E - Q \sigma) z -m_k^2 +
i\varrho }{Q + P z} \right) }{(z-T_1)(z-T_2)}dz, \\
 \mathcal{T}_0^+ &=& C_{\sigma}\int\limits_{0}^{\infty} \dfrac{  \ln\left(\pm \frac{(1-\beta\varphi)z +F}{\beta}\right) }{(z-T_1)(z-T_2)}dz,\\
 \mathcal{T}_0^- &=& C_{\sigma}\int\limits_{-\infty}^{0} \dfrac{  \ln\left(\pm \frac{(1-\beta\varphi)z +F}{\beta}\right) }{(z-T_1)(z-T_2)}dz.
\end{eqnarray}
We have already added to $D_0$ the extra terms which the sum of them is up to zero. 
These extra terms will contribute to the residue of $z$-poles when $\beta_{mlk}, \varphi_{mlk}$ 
become complex. 
\subsubsection{In the case of $D_{mlk}=0$}
In this case, the integrands of $D_{0}^{++}$, $D_{0}^{+-}$,$D_{0}^{-+}$ and $D_{0}^{--}$ have the form
\begin{eqnarray}
 \mathcal{I}(z,t)&=&\dfrac{1}{\Big[z+F_{nnlk}\Big]\Big[-2(\frac{A_{mlk}}{B_{mlk}}-\alpha_{lk})zt  
 - 2\frac{C_{mlk}}{B_{mlk}}t -2\frac{d_{lk}}{AC_{lk}}z +t^2-m_k^2+i\rho\Big]}\\
&=& \dfrac{1}{\Big[z+F_{nnlk}\Big]\Big[(K_{mlk}t+M_{mlk})z + L_{mlk}t +t^2-m_k^2+i\rho\Big]}.
\end{eqnarray}
Where new kinematic variables  are introduced
\begin{eqnarray}
 K_{mlk}&=&-2(\frac{A_{mlk}}{B_{mlk}}-\alpha_{lk}) \; \in \mathbb{R},\\
 L_{mlk}&=&-2\frac{C_{mlk}}{B_{mlk}}\; \in \mathbb{C},\\
 M_{mlk}&=&-2\frac{d_{lk}}{AC_{lk}}\; \in \mathbb{C}.
\end{eqnarray}
Instead of linearizing $t$, we are going to calculate the $z$-integrals
directly. The resulting reads
\begin{eqnarray}
 \int\limits_{0}^{\infty}dz \; \mathcal{I}(z,t)&=&\mathcal{H}(t)\Bigg[\ln(F_{nmlk})-\ln\Big(\frac{t^2+L_{mlk}t -m_k^2+i\rho}{K_{mlk}t+M_{mlk}}\Big) \Bigg],
 \\
\int\limits_{-\infty}^{0}dz \; \mathcal{I}(z,t)&=&-\mathcal{H}(t)\Bigg[\ln(-F_{nmlk})-\ln\Big(-\frac{t^2+L_{mlk}t -m_k^2+i\rho}{K_{mlk}t+M_{mlk}}\Big) \Bigg].
\end{eqnarray}
The $ \mathcal{H}(t)$ is defined as
\begin{eqnarray}
 \mathcal{H}(t)&=&\frac{1}{-t^2+(K_{mlk}F_{nmlk}-L_{mlk})t + M_{mlk}F_{nmlk} +m_k^2-i\rho}\nonumber\\
&=&-\frac{1}{(t-W_{nmlk}^{(1)})(t-W_{nmlk}^{(2)}) }. 
\end{eqnarray}
Where $ W_{nmlk}^{(1,2)}$ are given
\begin{eqnarray}
 W_{nmlk}^{(1,2)}=\dfrac{K_{mlk}F_{nmlk}-L_{mlk}\pm \sqrt{(K_{mlk}F_{nmlk}-L_{mlk})^2+4 (K_{mlk}F_{nmlk}+m_k^2-i\rho)}}{ 2}.
\end{eqnarray}
Finally, one gets

\begin{eqnarray}
\label{dzero}
 D_{0}^{++}&=&+i\pi^2\bigoplus_{nmlk}\int\limits_{0}^{\infty}dt f_{lk}g_{mlk} \mathcal{H}(t)\Bigg\{
   \ln(F_{nmlk}) - \ln\Big(\frac{t^2+L_{mlk}t -m_k^2+i\rho}{K_{mlk}t+M_{mlk}}\Big)  \Bigg\}, 
\\
D_{0}^{+-}&=&-i\pi^2\bigoplus_{nmlk}\int\limits^{0}_{-\infty}dt f_{lk}g_{mlk}^{-} \mathcal{H}(t)\Bigg\{
   \ln(F_{nmlk}) - \ln\Big(\frac{t^2+L_{mlk}t -m_k^2+i\rho}{K_{mlk}t+M_{mlk}}\Big)  \Bigg\},
\\
D_{0}^{-+}&=&+i\pi^2\bigoplus_{nmlk}\int\limits_{0}^{\infty}dt f_{lk}^{-}g_{mlk} \mathcal{H}(t)\Bigg\{
   \ln(-F_{nmlk}) - \ln\Big(-\frac{t^2+L_{mlk}t -m_k^2+i\rho}{K_{mlk}t+M_{mlk}}\Big)  \Bigg\}, \nonumber\\
 && \\
D_{0}^{--}&=&-i\pi^2\bigoplus_{nmlk}\int\limits^{0}_{-\infty}dt f_{lk}^{-}g_{mlk}^{-} \mathcal{H}(t)\Bigg\{
   \ln(-F_{nmlk}) - \ln\Big(-\frac{t^2+L_{mlk}t -m_k^2+i\rho}{K_{mlk}t+M_{mlk}}\Big)  \Bigg\}. \nonumber\\
\end{eqnarray}
It is easy to confirm that 
\begin{eqnarray}
\text{Im}\Big(L_{mlk}\Big)=
\begin{cases}
        \geq 0,   &\text{for} \; ++, -+,     \\\\
        \leq 0    &\text{for}\; +-, --,
\end{cases}
\quad \text{and} \quad \text{Im}\Big(M_{mlk}\Big)=
\begin{cases}
        \geq 0,   &\text{for}\;  ++, +-,    \\\\
        \leq 0    &\text{for}\;  -+,--.
\end{cases}
\end{eqnarray}
Moreover, one also verifies that 
\begin{eqnarray}
 \text{Im}\Big[t^2+L_{mlk}t -m_k^2+i\rho \Big]> 0 \quad \text{and}\quad  \text{Im} \Big[K_{mlk}t+M_{mlk}\Big]> 0. 
\end{eqnarray}
Noting that $X(t)$ is the second order polynomial written in terms of $t$ in the 
argument of logarithmic functions.
In detail,  it is
\begin{eqnarray}
X(t)=t^2+L_{mlk}t -m_k^2+i\rho=(t- X^{(1)}_{mlk})(t- X^{(2)}_{mlk}), 
\end{eqnarray}
with 
\begin{eqnarray}
 X^{(1,2)}_{mlk}=\dfrac{ -L_{mlk} \pm \sqrt{L_{mlk} ^2-4(-m_k^2+i\rho) }}{2}. 
\end{eqnarray}
Because $\text{Im}(X(t))$ and $\text{Im}(K_{mlk}t+M_{mlk})$ have the same sign, 
the logarithmic functions are decomposed as follows 
\begin{eqnarray}
 \ln\Big(\pm \frac{X(t)}{K_{mlk}t+M_{mlk}}\Big) &=& \ln [\pm X(t)] -\ln ( K_{mlk}t+M_{mlk}),       
 \end{eqnarray}
 and
 \begin{eqnarray}
 \ln[X(t)]&=&\ln(t- X^{(1)}_{mlk})+\ln(t- X^{(2)}_{mlk})+2\pi i \theta[\text{Im}(X^{(1)}_{mlk}) ]\theta[\text{Im}(X^{(2)}_{mlk}) ],           \\
\ln [-X(t)]&=&\ln(-t+ X^{(1)}_{mlk})+\ln(t- X^{(2)}_{mlk})-2\pi i \theta[\text{Im}(X^{(1)}_{mlk})]\theta[-\text{Im}(X^{(2)}_{mlk})].         
\end{eqnarray}
Finally, we arrive at
\begin{eqnarray}
 \dfrac{D_0}{ i \pi^2} &=& \bigoplus_{nmlk}\int\limits_{0}^{\infty}dt\; \mathcal{H}(t)\Bigg\{\;\; \chi_{nmlk}^{+}-f_{lk}g_{mlk} 
 \ln(t-X_{mlk}^{(1)})-f_{lk}^{-}g_{mlk} \ln(-t+X_{mlk}^{(1)})\nonumber\\
&&\hspace{0.8cm}-(f_{lk}g_{mlk} + f_{lk}^{-}g_{mlk}) \ln(t-X_{mlk}^{(2)})-(f_{lk}g_{mlk} + f_{lk}^{-}g_{mlk})
\ln(K_{mlk}t+M_{mlk})\nonumber \hspace{0.3cm}\Bigg\}\\
&&- \bigoplus_{nmlk} \int\limits^{0}_{-\infty}dt\; \mathcal{H}(t)\Bigg\{\;\; \chi_{nmlk}^{-}-f_{lk}g_{mlk}^{-}
\ln(t-X_{mlk}^{(1)})-f_{lk}^{-}g_{mlk}^{-} \ln(-t+X_{mlk}^{(1)})\nonumber\\
&&\hspace{0.8cm}-(f_{lk}g_{mlk}^{-} + f_{lk}^{-}g_{mlk}^{-}) \ln(t-X_{mlk}^{(2)})-(f_{lk}g_{mlk}^{-}
+ f_{lk}^{-}g_{mlk}^{-}) \ln(K_{mlk}t+M_{mlk})\nonumber \hspace{0.3cm}\Bigg\}.
\end{eqnarray}
Where the new kinematic variables are given
\begin{eqnarray}
 \chi_{nmlk}^{+}&=& f_{lk}g_{mlk}\ln(F_{nmlk})+ f_{lk}^{-}g_{mlk}\ln(-F_{nmlk})-2\pi i f_{lk}g_{mlk}
 \theta[\text{Im}(X_{mlk}^{(1)})]\theta[\text{Im}(X_{mlk}^{(2)})]\nonumber\\
&&+2\pi i f_{lk}^{-}g_{mlk}\theta[\text{Im}(X_{mlk}^{(1)})]\theta[-\text{Im}(X_{mlk}^{(2)})], \\
\chi_{nmlk}^{-}&=& f_{lk}g_{mlk}^{-}\ln(F_{nmlk})+ f_{lk}^{-}g_{mlk}^{-}\ln(-F_{nmlk}) 
- 2\pi i f_{lk}g_{mlk}^{-}\theta[\text{Im}(X_{mlk}^{(1)})]\theta[\text{Im}(X_{mlk}^{(2)})]\nonumber\\
&&+2\pi i f_{lk}^{-}g_{mlk}^{-}\theta[\text{Im}(X_{mlk}^{(1)})]\theta[-\text{Im}(X_{mlk}^{(2)})].
\end{eqnarray}
The remaining integrals (written in terms of $t$) will be integrated by using 
the basic integrals which are devoted in the Appendix.

In the next step, we will extend this work for evaluating tensor one-loop four-point 
functions with complex internal masses. In the parallel and orthogonal 
space~\cite{Kreimer:1991wj,Kreimer:1992ps,Bauer:2001ig}, 
a tensor one-loop $N$-point integral with rank $M$ can be decomposed as
\begin{eqnarray}
 T^{N}_{\mu_1\mu_2...\mu_M}=(-1)^{\frac{p_{\perp}}{2}}
\dfrac{\Big(g_{\mu_1\mu_2}...g_{\mu_{p_{\perp}}-1}g_{\mu_{p_{\perp}}}\Big)_{\text{sym}}}
{\mathcal{K}}\; T^{(p_0,p_1,...,p_{\perp})},
\end{eqnarray}
with 
\begin{eqnarray}
\mathcal{K} =
 \begin{cases}
          \prod\limits_{i=0}^{(p_{\perp}-2)/2}(n-J+2i),     & \text{if}   \;\; p_{\perp} \neq 0 ,\\
          1,     & \text{if}   \;\;p_{\perp}=0.
 \end{cases} 
\end{eqnarray}
Where space-time dimension is $n$ and $J$ is the number of parallel 
dimension (spanned by the external momenta). The tensor coefficients 
(form factors) are given
\begin{eqnarray}
		T^{p_0,p_1...p_{\bot}}_N&=&
		\frac{2\pi^{\frac{n-J}{2}}}{\Gamma(\frac{n-J}{2})}\int\limits_{-\infty}^{\infty} d l_{0}dl_{1}... dl_{J-1}\int\limits_{0}^{\infty} l_{\bot}^{n-J-1}d l_{\bot} \frac{l_0^{p_0}l_{1}^{p_1}...l_{J-1}^{p_{J-1}}l_{\bot}^{p_{\bot}}}{P_1P_2\cdots P_N}.
\end{eqnarray}

The traditional tensor reduction for one-loop integrals has been proposed by 
Passarino and Veltman~\cite{Passarino:1978jh}, later developed by Denner 
et al~\cite{Denner:2005nn}. In these schemes, 
the form factors will be obtained by contracting the Minkowski metric ($g_{\mu\nu}$) 
and external momenta into the tensor integrals. 
At this stage, we have to solve a system of linear equations where the Gram 
determinants appear in the denominator. If the Gram determinants will vanish 
or become very small, the reduction method will break or spoil numerical 
stability (this problem is called Gram determinant problem). The framework in this paper 
can be extended to calculate the form factors (or tensor one-loop integrals) directly. 
This will be devoted to future publication~\cite{khiemtensor}. It therefore opens a
new approach to solve Gram determinant problem analytically. 
\section{Numerical checks}
The calculation has been implemented into {\tt C$++$} program which is called
{\tt ONELOOP4PT.CPP}. In this program, the function uses the de facto input parameters 
of LoopsTools. The syntax of the new function is as follow
\begin{eqnarray}
 \text{ONELOOP4PT}(p_1^2, p_2^2, p_3^2, p_4^2, s,t, m_1^2, m_2^2, m_3^2, m_4^2, \rho),
\end{eqnarray}
with $s = (p_1+p_2)^2$ and $t = (p_2+p_3)^2$. 

In this section, we are going to check the program with {\tt LoopTools}
version $2.12$~\cite{Hahn:1998yk} (it is called {\tt LoopTools v.}$2.12$). 
In Tables $2$ and $3$, we check {\tt ONELOOP4PT.CPP} with {\tt LoopTools} in real 
and complex masses respectively. The input parameters are presented in these Tables.
One finds a good agreement between this work and {\tt LoopTools} in all cases. 
\begin{table}[ht]
\begin{center}
\begin{tabular}{|l|l|} \hline
$(p_1^{2}, p_2^2, p_3^2, p_4^2, s, t)\;$	& {\tt This work}  \\
                                        	& {\tt LoopTools v.$2.12$ }                          \\ \hline \hline
$(10,50,10,70, 170, 10)$  &$-1.2219797717173585\times 10^{-4}+2.0098337087139847\times 10^{-3}\; i$ \\
                                          &$-1.2219797696992298\times 10^{-4}+2.0098337092843695\times 10^{-3}\; i$\\ \hline
$(10,50,10,70, 170, -10)$ &$-1.4867162662896689\times 10^{-4}+1.6976243554156623\times 10^{-3}\; i$   \\
                                          & $-1.4867162664828184\times 10^{-4}+1.6976243552426918\times 10^{-3}\;i$\\ \hline
$(10,-50,10,-70, 170, 10)$ &$\;\;\;3.3519659312003411\times10^{-4}   + 2.9123620100294989\times 10^{-4}\; i$         \\
                                           &$\;\;\;3.3519659312270943\times 10^{-4}  + 2.9123620117879053\times 10^{-4}\; i$\\ \hline
\end{tabular}
\caption{\label{D05}In case of $(m_1^2,m_2^2,m_3^2,m_4^2)=(10, 20,30,40)$, and $ \rho=10^{-30}.$}
\end{center}
\end{table}
\begin{table}[ht]
\begin{center}
\begin{tabular}{|l|l|} \hline
$(p_1^{2}, p_2^2, p_3^2, p_4^2, s, t)$	&  ({\tt This work})$\times 10^{-4}$ \\ 
& ({\tt LoopTools v.$2.12$})$\times 10^{-4}$ \\ \hline \hline
$(10,60,10,90,200,10)$ & $-7.6754958275901917+  7.93692363083359122\; i$\\
                                       & $ -7.6754958275902069+  7.93692363083357794\; i $ \\ \hline
$(10,60,-10,90,200,10)$ &$  -6.2213615288278696+  8.20289788323527371\; i$\\
                                        & $ -6.2213615288278837 +  8.20289788323525900\; i$\\ \hline
$(10,-60,-10,-90,200,-10)$ & $\;\;\;1.5318455243668001 + 2.56186787016916487\; i$\\
                                           & $\;\;\;1.5318455243667963 + 2.56186787016915870\; i$\\ \hline
$(10,60,0,0,200,-10)$        &$-3.3499315746623337 + 6.17862189272097645\; i$ \\
                                           &$-3.3499315746623217 + 6.17862189272098160\; i$\\ \hline
\end{tabular}
\caption{\label{D08}In case of $m_1^2=10-5i,m_2^2=20-2i, m_3^2=30-3i, m_4^2=40-4i$, and $ \rho=10^{-30}.$}
\end{center}
\end{table}

In Table \ref{D06}, we compare the results generated by 
{\tt ONELOOP4PT.CPP} with {\tt LoopTools} by changing the value 
of $m_3^2$. Other input parameters are fixed  as follows
$(p_1^{2}, p_2^2, p_3^2, p_4^2, s, t)=(10, -60,-10,-90,200,-10)$ 
and $(m_1^2,m_2^2,m_4^2)=(10-5i,20-2i,40-4i)$, and $ \rho=10^{-30}.$
One again finds a good agreement between the results computed from
this work and {\tt LoopTools} in all cases of $m_3^2$. 

\begin{table}[ht]
\begin{center}
\begin{tabular}{|c|l|} \hline		
$m_3^{2}$ & {\tt This work} \\ 
               & {\tt LoopTools v.$2.12$}\\ \hline \hline
$10-3i$   &$ 2.2251608819818614\times 10^{-4}+3.6032814993796746\times 10^{-4}\; i$     \\
               & $2.2251608819818662\times 10^{-4}+3.6032814993796724\times 10^{-4}\; i$ \\ \hline
$100-3i$ &$ 8.3454216851333892\times 10^{-5}+1.3928616354428907\times 10^{-4}\; i$ \\
               &$ 8.3454216851334176\times 10^{-5}+1.3928616354428922\times 10^{-4}\; i$  \\\hline
$1000 -3i$ & $1.5581298826002613\times 10^{-5}+2.2404821779612569\times 10^{-5}\; i$\\
                  & $1.5581298826002624\times 10^{-5}+2.2404821779612618\times 10^{-5}\;i$\\ \hline
$100000 -3i$ & $1.8308170810361346\times 10^{-6}+2.4152459663910780\times 10^{-6}\; i$ \\
                      & $1.8308170810361369\times 10^{-6}+2.4152459663910805\times 10^{-6}\; i$ \\ \hline
\end{tabular}
\caption{\label{D06}In case of $(p_1^{2}, p_2^2, p_3^2, p_4^2, s, t)=(10, -60,-10,-90,200,-10)$ 
and $(m_1^2,m_2^2,m_4^2)=(10-5i,20-2i,40-4i)$, and $ \rho=10^{-30}.$}
\end{center}
\end{table}
\section{Conclusions} 
\noindent
In this paper, we have presented the analytic solution for scalar one-loop four-point 
integrals with real and complex internal masses. This method can be extended to calculate tensor integrals directly. 
It may open a new way to cure the inverse determinant problem analytically.  
In the numerical checks, one compared this work with {\tt LoopTools}. We found a good agreement between the 
results generated from this work and  the ones from {\tt LoopTools}. In future work, we will 
proceed this method for evaluating tensor one-loop four-point integrals.\\

{\bf Acknowledgment:}~
This research is funded by Vietnam National Foundation for Science and Technology Development (NAFOSTED) under grant number 103.01-2016.33. The author is also grateful to Chau Thien Nhan for reading the manuscript and to all members of Theoretical Physics Department, University of Science HoChiMinh City for fruitful discussions.
K. H. Phan is grateful to  Dr. Do Hoang Son
for fruitful discussions and his contribution to this work. 
\section*{Appendix}
In this Appendix, we present several useful formulae. The first one we mention 
\begin{eqnarray}
\label{logdecompose}
 \ln(a b)&=&\ln (a)+\ln (b)+\eta(a,b),\\
 \ln\left( \frac{a}{b} \right)&=&\ln (a)-\ln (b)+\eta\left( a,\frac{1}{b} \right).
\end{eqnarray}
Where the Eta function is defined as
\begin{eqnarray}
\eta(a,  b)&=&2\pi i\Bigg\{\theta\Big[-\text{Im}(a)\Big]\theta\Big[-\text{Im}(b)\Big]\theta\Big[\text{Im}(ab)\Big]
-\theta\Big[\text{Im}(a)\Big]\theta\Big[\text{Im}(b)\Big]\theta\Big[-\text{Im}(ab)\Big]\Bigg\}. \nonumber\\
\end{eqnarray}

Let us consider $f(z)$ is a rational function with $\lim\limits_{z\rightarrow \infty}f(z)=0$ and  
no poles on the negative real $z$-axes, $z_k$ are poles of $f(z)$. By closing the integration contour as 
Fig.~\ref{closewick}, one then can derive the following relation
\begin{eqnarray}
\oint f(z) \ln(z)\; dz &=& 2 i \pi \sum_k \mathcal{R}\text{es}\{ f(z) \ln(z); z_k\}     \\
&=& \left\{\int\limits^{0}_{-\infty} + \int\limits_0^{-\infty} + \int\limits_{\Gamma_k} + \int\limits_{C_k}\right\} f(z)
\ln(z) dz \\
&=& \int\limits^{0}_{-\infty} f(x) \ln(x) dx + \int\limits_0^{-\infty} f(x e^{-2 i \pi}) (\ln(x) - 2 i
\pi) dx \\
&=& 2 i\pi \int\limits^{0}_{-\infty} f(z) dz . 
\end{eqnarray}
We then arrive at  the relation 
\begin{eqnarray}
\label{masterLog}
\int\limits_{-\infty}^a f(z) dz &=& + \sum_k \mathcal{R}\text{es}\{ f(z) \ln(z -a); z_k\} = \int\limits_{-a}^{\infty} f(-z)dz.
\end{eqnarray}
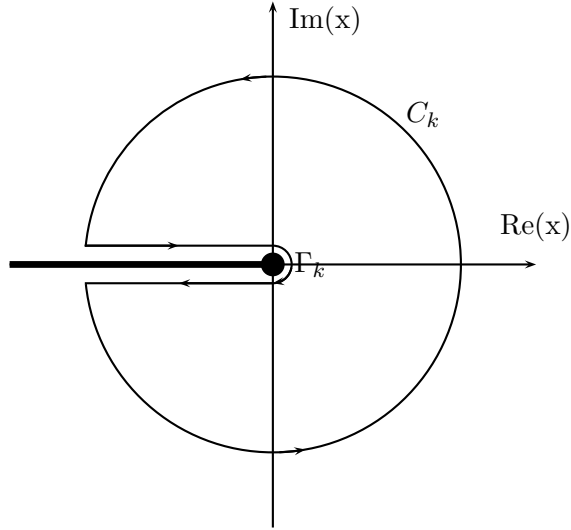
\begin{figure}[ht]
\begin{center}
\begin{pspicture}(-6, -3)(6, 3)
\psline(-2.5, 0.25)(0, 0.25)
\psline{->}(-2.5, 0.25)(-1.25, 0.25)
\psline(-2.5, -0.25)(0, -0.25)
\psline{<-}(-1.25, -0.25)(0, -0.25)
\psarc{->}(0, 0){2.5}{0}{100}
\psarc(0, 0){2.5}{92}{174.5}
\psarc{->}(0, 0){2.5}{186}{280}
\psarc(0, 0){2.5}{272}{360}
\psarc{<-}(0, 0){0.25}{-90}{-10}
\psarc(0, 0){0.25}{-90}{90}
\psline{->}(-3.5, 0)(3.5,0)
\psline{->}(0, -3.5)(0, 3.5)
\psset{dotsize=9pt 0}
\psdots(0,0)
\rput(2,2){$C_{k}$}
\rput(3.5,0.5){Re(x)}
\rput(0.7,3.25){Im(x)}
\rput(0.5,0){$\Gamma_{k}$}
 \psline[linewidth=0.09cm](-3.5, 0)(0, 0)
\end{pspicture}
\end{center}
\caption{\label{closewick} The integration contour.}
\end{figure}

We consider three basic integrals in the following paragraphs.  
\begin{enumerate}
 \item \underline{Basic integral $I$:}\\
 
The basics integral $I$ is defined as
\begin{eqnarray}
\mathcal{R}_1(x,y)= \int\limits_{0}^{\infty}\frac{1}{(z+x)(z+y)}dz=\frac{\ln(x)-\ln(y)}{x-y},
\end{eqnarray}
with $x,y\in \mathbb{C}$.\\\\

\item \underline{Basic integral $II$:} \\

The basic integral $II$ is
\begin{eqnarray}
\mathcal{R}_2(r,x,y) = \int\limits_{0}^{\infty}\frac{\ln(1+rz)}{(z+x)(z+y)}dz 
&=&-\frac{1}{x-y}\Big[ \text{Li}_2(1-rx)-\text{Li}_2(1-ry) \Big]\\ \nonumber&&-\frac{1}{x-y}
\Big[  \eta(x,r)\ln(1-rx)  -\eta(y,r)\ln(1-ry)     \Big],
\end{eqnarray}
with $r , x,y\in \mathbb{C}$.

\item \underline{Basic integral $III$:} \\

The basic integral $III$ has the form of
\begin{eqnarray}
 \mathcal{R}_3 &=& \int\limits^{\infty}_{-\infty} dz\; G(z)\theta\Big[\text{Im}
 \left(\frac{ S(\sigma,z)}{Pz+Q} \right) \Big],
\end{eqnarray}
with $G(z)$ and $S(\sigma,z) $ are defined in Eqs.~(\ref{GZ}, \ref{SZ}) respectively. \
We have known that $\text{Im} \Big(\frac{ S(\sigma,z)}{Pz+Q}\Big)$ is independent 
of $\sigma$,  we then can expand the integrand as
\begin{eqnarray}
\int\limits^{\infty}_{-\infty} dz\; G(z)\theta\Big[\text{Im}\left( \frac{ S(\sigma,z)}{Pz+Q} \right) \Big]
&=&  \int\limits^{\infty}_{-\infty} dz\; G(z)\theta\Big[A_0 z^2+B_0z+C_0\Big].
\end{eqnarray}
Where $A_0, B_0, C_0$ are given by
\begin{eqnarray}
 A_0&=&P\;\text{Im}(E), \\
 B_0&=&P\Gamma_k+\rho\;P+\text{Re(Q)}\text{Im}(E)-\text{Im}(Q)\text{Re}(E), \\
 C_0&=&\text{Im}(Q)\text{Re}(m_k^2)+\text{Re}(Q)(\Gamma_k+\rho)
\end{eqnarray}
For $ \text{Im}\Bigg(\dfrac{ S(\sigma,z)}{Pz_0+Q}\Bigg) \geq 0$\
in the region $\Omega \subset \mathbb{R}$, one then has
\begin{eqnarray}
  \int\limits^{\infty}_{-\infty} dz\; G(z)\theta\Big[ \text{Im} 
  \left( \frac{ S(\sigma,z)}{Pz+Q} \right)\Big]&=& \int\limits_{\Omega} dz\; G(z).\nonumber
\end{eqnarray}
The integral in right-hand side of this equation is nothing but it will be reduced to basic integral $I$.
\end{enumerate}


\begin{thebibliography}{100}
\bibitem{ATLAS:2013hta}
ATLAS Collaboration,
  arXiv:1307.7292 [hep-ex].
\bibitem{CMS:2013xfa}
  CMS Collaboration,
  arXiv:1307.7135.
%
\bibitem{Baer:2013cma}
  H.~Baer et al., 
  arXiv:1306.6352 [hep-ph].
\bibitem{Passarino:1978jh}
  G.~Passarino and M.~J.~G.~Veltman,
  Nucl.\ Phys.\ B {\bf 160} (1979) 151.
  doi:10.1016/0550-3213(79)90234-7
\bibitem{Denner:2005nn}
  A.~Denner and S.~Dittmaier,
  Nucl.\ Phys.\ B {\bf 734} (2006) 62
  doi:10.1016/j.nuclphysb.2005.11.007
  [hep-ph/0509141].
\bibitem{Denner:2005fg}
  A.~Denner, S.~Dittmaier, M.~Roth and L.~H.~Wieders,
  Nucl.\ Phys.\  B {\bf 724} (2005) 247
  [hep-ph/0505042].
\bibitem{'tHooft:1978xw}
  G.~'t Hooft and M.~J.~G.~Veltman,
  Nucl.\ Phys.\ B {\bf 153} (1979) 365.
%
\bibitem{Denner:1991qq}
  A.~Denner, U.~Nierste and R.~Scharf,
  Nucl.\ Phys.\ B {\bf 367} (1991) 637.
\bibitem{Nhung:2009pm}
  D.~T.~Nhung and L.~D.~Ninh,
  Comput.\ Phys.\ Commun.\  {\bf 180} (2009) 2258
  [arXiv:0902.0325 [hep-ph]].
\bibitem{Denner:2010tr}
  A.~Denner and S.~Dittmaier,
  Nucl.\ Phys.\ B {\bf 844} (2011) 199
  doi:10.1016/j.nuclphysb.2010.11.002
  [arXiv:1005.2076 [hep-ph]].
  
\bibitem{Kreimer:1991wj}
  D.~Kreimer,
  Z.\ Phys.\ C {\bf 54} (1992) 667.
  doi:10.1007/BF01559496
\bibitem{Kreimer:1992ps}
  D.~Kreimer,
  Int.\ J.\ Mod.\ Phys.\ A {\bf 8} (1993) 1797.
  doi:10.1142/S0217751X93000758
\bibitem{Franzkowski}
J.~Franzkowski, {\it Dissertation}, Mainz 1997.
\bibitem{vanOldenborgh:1989wn}
  G.~J.~van Oldenborgh and J.~A.~M.~Vermaseren,
  Z.\ Phys.\ C {\bf 46} (1990) 425.
\bibitem{Ellis:2007qk}
  R.~K.~Ellis and G.~Zanderighi,
  JHEP {\bf 0802} (2008) 002
  doi:10.1088/1126-6708/2008/02/002
  [arXiv:0712.1851 [hep-ph]].

\bibitem{vanHameren:2010cp}
  A.~van Hameren,
  Comput.\ Phys.\ Commun.\  {\bf 182} (2011) 2427
  doi:10.1016/j.cpc.2011.06.011
  [arXiv:1007.4716 [hep-ph]].

\bibitem{Binoth:2008uq}
  T.~Binoth, J.-P.~Guillet, G.~Heinrich, E.~Pilon and T.~Reiter,
  Comput.\ Phys.\ Commun.\  {\bf 180} (2009) 2317
  doi:10.1016/j.cpc.2009.06.024
  [arXiv:0810.0992 [hep-ph]].
\bibitem{Cullen:2011kv}
  G.~Cullen, J.~P.~Guillet, G.~Heinrich, T.~Kleinschmidt, E.~Pilon, T.~Reiter and M.~Rodgers,
  Comput.\ Phys.\ Commun.\  {\bf 182} (2011) 2276
  doi:10.1016/j.cpc.2011.05.015
  [arXiv:1101.5595 [hep-ph]].
\bibitem{Guillet:2013msa}
  J.~P.~Guillet, G.~Heinrich and J.~F.~von Soden-Fraunhofen,
  Comput.\ Phys.\ Commun.\  {\bf 185} (2014) 1828
  doi:10.1016/j.cpc.2014.03.009
  [arXiv:1312.3887 [hep-ph]].
\bibitem{Bern:1992em}
  Z.~Bern, L.~J.~Dixon and D.~A.~Kosower,
  Phys.\ Lett.\ B {\bf 302} (1993) 299
   Erratum: [Phys.\ Lett.\ B {\bf 318} (1993) 649]
  doi:10.1016/0370-2693(93)90469-X, 10.1016/0370-2693(93)90400-C
  [hep-ph/9212308].
\bibitem{Duplancic:2000sk}
  G.~Duplancic and B.~Nizic,
  Eur.\ Phys.\ J.\ C {\bf 20} (2001) 357
  doi:10.1007/s100520100675
  [hep-ph/0006249].
\bibitem{Hahn:1998yk}
  T.~Hahn and M.~Perez-Victoria,
  Comput.\ Phys.\ Commun.\  {\bf 118} (1999) 153
  doi:10.1016/S0010-4655(98)00173-8
  [hep-ph/9807565].
, 
\bibitem{Bauer:2001ig}
  C.~Bauer and H.~S.~Do,
  Comput.\ Phys.\ Commun.\  {\bf 144} (2002) 154
  doi:10.1016/S0010-4655(02)00158-3
  [hep-ph/0102231].
  
\bibitem{Berger:2008ag}
  C.~F.~Berger, Z.~Bern, L.~J.~Dixon, F.~Febres Cordero, D.~Forde, H.~Ita, D.~A.~Kosower and D.~Maitre,
  Nucl.\ Phys.\ Proc.\ Suppl.\  {\bf 183} (2008) 313
  doi:10.1016/j.nuclphysbps.2008.09.123
  [arXiv:0807.3705 [hep-ph]].
\bibitem{Ossola:2007ax}
  G.~Ossola, C.~G.~Papadopoulos and R.~Pittau,
  JHEP {\bf 0803} (2008) 042
  doi:10.1088/1126-6708/2008/03/042
  [arXiv:0711.3596 [hep-ph]].
\bibitem{Carrazza:2016gav}
  S.~Carrazza, R.~K.~Ellis and G.~Zanderighi,
  Comput.\ Phys.\ Commun.\  {\bf 209} (2016) 134
  doi:10.1016/j.cpc.2016.07.033
  [arXiv:1605.03181 [hep-ph]].
\bibitem{Actis:2016mpe}
  S.~Actis, A.~Denner, L.~Hofer, J.~N.~Lang, A.~Scharf and S.~Uccirati,
  arXiv:1605.01090 [hep-ph].
\bibitem{Denner:2016kdg}
  A.~Denner, S.~Dittmaier and L.~Hofer,
  Comput.\ Phys.\ Commun.\  {\bf 212} (2017) 220
  doi:10.1016/j.cpc.2016.10.013
  [arXiv:1604.06792 [hep-ph]].
\bibitem{Heinrich:2008si}
  G.~Heinrich,
  Int.\ J.\ Mod.\ Phys.\ A {\bf 23} (2008) 1457
  doi:10.1142/S0217751X08040263
  [arXiv:0803.4177 [hep-ph]].
\bibitem{Borowka:2012yc}
  S.~Borowka, J.~Carter and G.~Heinrich,
  Comput.\ Phys.\ Commun.\  {\bf 184} (2013) 396
  doi:10.1016/j.cpc.2012.09.020
  [arXiv:1204.4152 [hep-ph]].
\bibitem{Borowka:2015mxa}
  S.~Borowka, G.~Heinrich, S.~P.~Jones, M.~Kerner, J.~Schlenk and T.~Zirke,
  Comput.\ Phys.\ Commun.\  {\bf 196} (2015) 470
  doi:10.1016/j.cpc.2015.05.022
  [arXiv:1502.06595 [hep-ph]].
\bibitem{Borowka:2017idc}
  S.~Borowka, G.~Heinrich, S.~Jahn, S.~P.~Jones, M.~Kerner, J.~Schlenk and T.~Zirke,
  arXiv:1703.09692 [hep-ph].
\bibitem{Yuasa:2011kt}
  F.~Yuasa, T.~Ishikawa, Y.~Kurihara, J.~Fujimoto, Y.~Shimizu, N.~Hamaguchi, E.~de Doncker and K.~Kato,
  PoS CPP {\bf 2010} (2010) 017
  [arXiv:1109.4213 [hep-ph]].
\bibitem{Gluza:2007rt}
  J.~Gluza, K.~Kajda and T.~Riemann,
  Comput.\ Phys.\ Commun.\  {\bf 177} (2007) 879
  doi:10.1016/j.cpc.2007.07.001
  [arXiv:0704.2423 [hep-ph]].
\bibitem{khiemtensor}
 Khiem Hong Phan, {\it in preparation}. 
\end{thebibliography}
\end{document}